\documentclass[aps,prx,twocolumn,groupedaddress,amsmath,amssymb,showpacs]{revtex4-2}

\usepackage{color,soul}
\usepackage{graphicx,graphics,color,epsfig}
\usepackage{amsmath,amssymb,amsfonts,amsfonts}
\usepackage{epstopdf}
\usepackage{bm}
\usepackage{booktabs}
\usepackage{times,xspace}
\usepackage[utf8]{inputenc}
\usepackage{hyperref}
\usepackage{float}
\usepackage[normalem]{ulem}
\usepackage{filecontents}
\usepackage[normalem]{ulem}

\def\be{\begin{equation}}
\def\ee{\end{equation}}
\def\bea{\begin{eqnarray}}
\def\eea{\end{eqnarray}}

\def\PT{$\mathcal{PT}$}

\begin{document}

\title{A Hermitian bypass to the non-Hermitian quantum theory}

\author{Priyanshi Bhasin} 
\author{Tanmoy Das}
\email{tnmydas@iisc.ac.in}
\affiliation{Department of Physics, Indian Institute of Science, Bangalore 560012, India.}

\date{\today}
\vspace{0.3cm}

\begin{abstract}
Describing systems with non-Hermitian (NH) operators remains a challenge in quantum theory due to instabilities (e.g., exceptional points and decoherence) arising from interactions with the environment. 
We propose a framework to express the energy states of NH Hamiltonians using a well-defined basis (dub \textit{computational} basis) derived from a related Hermitian operator. This suitably shifts the  singularities from the basis states to the expansion coefficients, allowing for their easier mathematical treatment and parametric control. Furthermore, we introduce a `space-time' transformation on the computational basis that defines a generic dual space map for the energy states. Interestingly, this transformation leads to a symmetry for real/imaginary energy values, revealing the existence of weaker condition than hermiticity or the \PT~symmetry. 
This leads to clearer understanding and novel interpretations of key features like exceptional points, dual space, and weaker symmetry-enforced real eigenvalues. The applicability of our framework extends to various branches of physics where NH operators manifest as ladder operators, order parameters, self-energies, projectors, and other entities.
\end{abstract}
\maketitle

\section{Introduction}\label{Sec:Introduction}
Non-Hermitian (NH) operators are more common than we often realize across various domains. The presence of NH operators extends beyond their relevance in open quantum systems, non-equilibrium states, and quantum optical devices. 
In closed quantum systems, they also appear in disguise, such as the  NH bosonic Bogoliubov-Valatin Hamiltonian~\cite{BogoluvbovValatin, BogoluvbovValatin_2, BdG2NH}, effective NH Hamiltonian acting on a sub-Hilbert space or sub-systems~\cite{WavefuncNH, NHQM, QScarBerry, QScarReview, QFragmentation}, chiral boundary Hamiltonians of topological insulators~\cite{VishwanathNH, SatoSkinEffect}, complex order parameters for phase transitions, coherent states of ladder operators, self-energy dressed Hamiltonians in many-body quantum theory~\cite{SelfenergyEmil,SelfenergyNagaosa}, auxiliary boundary conditions or degrees of freedom in numerical  simulations~\cite{NHQSBook}, and others. In such cases, we grapple with the theory of NH operators which lacks a consistent and unified framework.

The quantum theory of NH operators deviates significantly from the Hermitian quantum framework. For instance, the NH theory encounters singularities such as exceptional points~\cite{Heiss_2012_Review}, indefinite inner products~\cite{Bender_2015}, and other instabilities originating from interactions with the environment. 
For example, contrasted with the Hermitian quantum theory, for NH quantum theory with biorthonormal basis vectors, there exists no generic map to the dual space. Consequently,  physical observables lack unique definitions. Unitary operators cannot define gauge redundancy and gauge symmetry. The eigenvectors belong to a Hilbert space with indefinite metric. The time derivative does not simply yield a commutation relation with the NH Hamiltonian (\textit{c.f.} Heisenberg equation of motion), the time-evolution is non-unitary, and so on. 

Is it possible to establish a unique and consistent quantum theory for general NH operators? Such a theory would simplify the study of open quantum systems, including jumpfree as well as a wider class of systems involving quantum jumps~\cite{JumpfreePhysRevA, JumpPerturbationPhysRevA}, which are often approached through master equations. The master equation formalism involves density operators to capture the dissipative effects of interaction with the environment whereas the effective NH quantum Hamiltonian approach captures these effects through the evolution of a pure state. Notably, this offers a significant advantage, both in analytical~\cite{JumpfreePhysRevA, JumpPerturbationPhysRevA} and numerical considerations~\cite{Molmer:93}, in view of the fact that the density matrix involves $n^2$ variables for an $n$-dimensional Hilbert as opposed to the wavefunction involving $\sim n$ variables. These considerations also find application in other relevant studies, for instance, in providing an alternative perspective on von Neumann measurement scheme~\cite{vonNeumannNHIOP}, etc. However, due to the features of non-Hermiticity listed in the previous paragraph, the diagonalizabiltiy of NH Hamiltonians is `delicate'~\cite{JumpfreePhysRevA} and other features, such as dual space maps, etc., lack a complete framework and consistent interpretation. The objective of this work is to make significant strides in developing the quantum theory of general NH operators, thereby setting the stage for studying their dynamics.

Here we develop a formalism for studying NH Hamiltonians by defining a (real-valued) scaling parameter as a `constant of motion' and mapping the problem to this curve of `constant of motion'. The `constant of motion' is determined by the eigenvalue of the (Hermitian) anti-commutator of $H$ and $H^{\dagger}$; thereby, it establishes a bridge between $H$ and $H^{\dagger}$, giving rise to a symmetry operator and dual space mapping. Taking advantage of this relation, we consider a Hermitian operator $F:=H^{\dagger}H$. Moiseyev~\cite{NHQM,Moiseyev2} showed that a related Hermitian operator can be used to find bounds on energies for NH Hamiltonians but did not use the eigenstates of the constructed Hermitian operator for any exact analysis of the complex energies and eigenvectors, or to achieve the dual space maps, weaker symmetry constraints, etc. Feinberg and Zee~\cite{JoshuaZee}, introduced the method of `Hermitization' where they considered a chiral operator in an extended Hilbert space (refer Appendix~\ref{Appendix:Chiral}) as an auxiliary Hermitian operator to analyze its component non-Hermitian Hamiltonian ($H$) of interest.  This method, while useful in the random NH matrix theory, did not solve for the energy eigenvalues and eigenvectors. Their chiral Hermitian operator also fails to capture all the distinct symmetry classes of the NH Hamiltonian~\cite{Bernard_LeClair_2002}.

In the present work, by choosing the eigenvectors of $F$ as a basis,  namely the \textit{computational basis}, we establish a framework that offers various advantages, including simplifications and mathematical rigor, and also captures the symmtery classes of the NH Hamiltonian appropriately. The hermiticity of $F$ ensures the computational basis is well defined, i.e., free from exceptional points, or other singularities. The eigenvalues of  $F$ are found to belong to a continuous and bounded interval whose \textit{bounds are} the \textit{exceptional points} (or contours) and the \textit{centre is the normal point} where $H$ and $H^{\dagger}$ commute. The $H$ serves as a ladder operator in the computational basis, and the `annihilation' of the basis states at the two boundaries of the interval turned out to be the  `exceptional points', i.e., they serve as `vacua'. 

By expressing the Hamiltonian eigenstates in this computational basis, we further discover several properties. (a) The energy eigenstates \textit{coalesce} into one of the computational states at exceptional points. (b) The strength of the "non-Hermitian effects" is quantified by a real parameter which is precisely an eigenvalue of the Hermitian operator $F$. (c) The dual space can be \textit{(anti-/) linearly mapped} via a discrete `space-time' transformation, albeit with a negative norm, suggesting an indefinite inner product space. (d) The dual space map is dynamic, in general, involving system-environment coupling but leads to a symmetry for real/imaginary energy eigenvalues. This implies that there exists a weaker condition on real energies than the hermiticity or the \PT-symmetry. 
We discuss appropriate examples that unravel the underlying interpretation of exceptional points, normal operators, topology, dual space, weaker symmetry-enforced real eigenvalues, and others. We extend the theory to higher dimensional Hilbert spaces and consequently predict that in odd-dimension, at least one complex energy eigenvalue exists whose absolute value becomes parameter-free (dub `flat-energy' or `flat band'). Finally, we discuss the potential applications of our formalism in various contexts.

The rest of the paper is structured as follows. In Sec. \ref{Sec:Comp basis}, we introduce the core concepts and the essential components of our formalism. Here, we construct the operator $H^{\dagger}H$, the computational basis, the ladder operations, and exceptional points as vacua. Sec. \ref{Sec:Maps Dual Sym} is dedicated to the development of dual space maps as dynamical transformations and the inherent symmetry in the NH Hamiltonians. We apply our formalism to a selection of representative examples in Sec. \ref{Sec:App Example}. In Sec. \ref{Sec:Hig dim}, we extend the results to higher dimensional Hilbert spaces and discuss two interesting special cases of degeneracies, namely the \textit{circular} and \textit{point degeneracies}. We summarize our work in Sec. \ref{Sec:Discussion} and conclude with a discussion on the applicability of our formalism in diverse settings and outline the remaining works in this direction.

\section{The Computational Basis} \label{Sec:Comp basis}

\begin{figure}[t]
  \centering  \includegraphics[width=8.6cm]{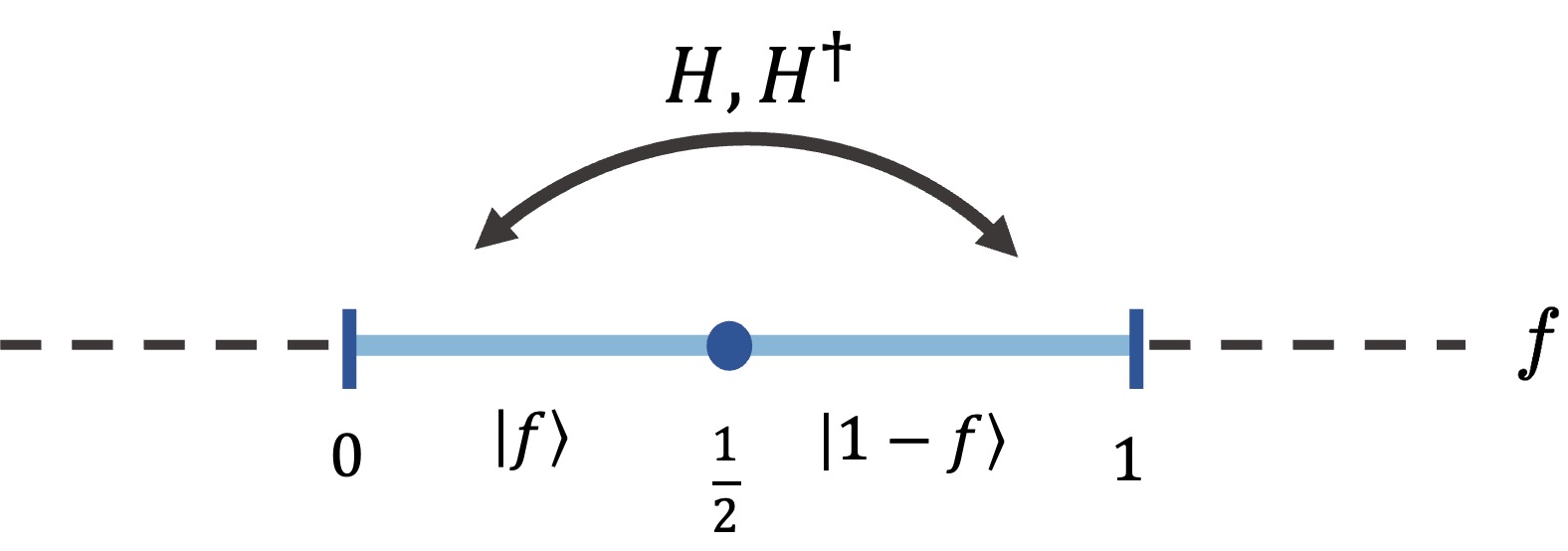}
  \caption{The span of the computational eigenvalue space between the two exceptional points/contours at $f=0, 1$, and a normal point at $f=1/2$, and the cyclic ladder actions of $H$ and $H^\dagger$ across the normal point are schematically shown here. }
  \label{Fig:Ladder action}
\end{figure}

For any general NH Hamiltonian $H$, we can construct a positive Hermitian operator $D$ as follows $\{ H,H^\dagger \}:=D$. In this study, we focus on the NH systems for which the operator $D$ is a scalar multiple of the identity operator, i.e, $d\mathbb{I}$. As will be shown in the subsequent sections, this class encapsulates general NH Hamiltonians for Hilbert space dimension equal to two. However, it weakly violates generality for Hilbert space dimensions more than two; we will see further in Sec. \ref{Sec:Hig dim}. But the formalism can be made  general by taking into account all eigenvalues of $D$. We will consider such a case in future studies but focus here on a single eigenvalue $d$.

Without loss of generality, we can either set $d=1$ and effectively treat $d$ as a 'constant of motion', or rescale the Hamiltonian to be \textit{dimensionless} as $H\rightarrow H/\sqrt{d}$, such that we obtain an anticommutation algebra: 
\begin{equation}\label{Eq:h hdag one}
   \left\{ H,H^\dagger \right\} = \mathbb{I},
\end{equation}
where $\mathbb{I}$ is the unit matrix. Hence, this construction relates $H$ and $H^\dagger$ through the anticommutation which we exploit in Sec.~\ref{Sec:Maps Dual Sym} to construct the dual space maps. (See Supplemental Material~\cite{Supp} for generalizations to the $d\neq 1$ case)

Next, we attempt to construct a Hermitian operator in whose basis
the above algebra in Eq.~\eqref{Eq:h hdag one} helps $H$ and $H^\dagger$ serve as ladder operators. A candidate choice is:
\begin{equation}
  F:=H^\dagger H.
  \label{Eq:f def}
\end{equation}
$F$ is a positive, Hermitian operator with a physical dimension of $[E^2]$. Suitably, $F$ inherits the information of both $H$ and $H^{\dagger}$, but the phase of the complex energy is dropped out. The eigenstates of $F$ are orthonormal: 
\begin{align}\label{Eq:f eval}
    F|f\rangle = f|f\rangle,
\end{align}
where $f\in \mathbb{R}$ is bounded from below as $f\ge 0$. $|f\rangle$ states are orthonormalized as $\langle f|f'\rangle=\delta_{f,f'}$. Using the algebra in Eq.~\eqref{Eq:h hdag one}, we infer two important properties:

\begin{enumerate}
    \item \textit{The eigenvalues of $F$ belong to a continuous and bounded interval.}
    
    $f$ takes continuous values but bounded from both sides (since $F$ is positive and  Hermitian) as   
    \begin{align}
        0 \leq f \leq 1.
        \label{Eq:f bound}
    \end{align}
 The span of $f$ has three special points: the two boundaries (vacua) $f=0$, $1$ correspond to the \textit{exceptional points} of $H$, while $f=1/2$ corresponds to a degenerate point for $F$ where $H$ becomes a \textit{normal operator} (see Fig.~\ref{Fig:Ladder action}). We elaborate on these points in Sec. \ref{Subsec:EP NP}. For future convenience, we  partition this interval into two halves: $\mathcal{F}_{0 \leqslant f \leqslant\frac{1}{2}}$ and $\mathcal{F}_{\frac{1}{2} \leqslant f \leqslant 1}.$
   
    \item \textit{$H$ and $H^{\dagger}$ are the ladder operators.} 
    
    $H |f\rangle$ and $H^{\dagger}|f\rangle$ are also eigenstates of $F$ with the \textit{same} eigenvalue of $(1-f)$. So both $H$ and $H^{\dagger}$ act as ladder operators between $\mathcal{F}_{0 \leqslant f \leqslant\frac{1}{2}}$ and $\mathcal{F}_{\frac{1}{2} \leqslant f \leqslant 1}$. We consider here the non-degenerate case, while the degenerate case will be studied in Appendix \ref{Appendix:NP in 2D}. The magnitudes of proportionality constants can be easily evaluated, and we consider two arbitrary phases $\phi$ and $\gamma$ in the ladder action as
\begin{subequations}\label{Eq:ladder action}
\begin{align}
H|f\rangle &= e^{i(\gamma+\phi)} \sqrt{f}|1-f\rangle, \label{Eq:ladder action a}\\
H|1-f\rangle &= e^{i(\gamma-\phi)} \sqrt{1-f}|f\rangle  \label{Eq:ladder action b},\\
H^{\dagger}|f\rangle &= e^{-i(\gamma-\phi)} \sqrt{1-f}|1-f\rangle, \label{Eq:ladder actionc}\\
H^{\dagger}|1-f\rangle &= e^{-i(\gamma+\phi)} \sqrt{f}|f\rangle. \label{Eq:ladder action d}
\end{align}
\end{subequations}
Such phase factors in the ladder operations are generally absent in the standard Fock space construction of the non-Abelian group or the Harmonic oscillator case. However, we discover below that the two phases are related to the complex phase of the eigenvalues and eigenstates of $H$. 
\end{enumerate}  

\subsection{Eigenspectrum of $H$}\label{Subsec:H eigenspec}
For a two-dimensional Hamiltonian (higher dimension is considered in Sec. \ref{Sec:Hig dim}), we can now represent the NH $H$ in the well-defined computational basis of $F$ as
\begin{align}
H=\sqrt{f}e^{i(\gamma+\phi)}|1-f\rangle\langle f| +\sqrt{1-f}e^{i(\gamma-\phi)}|f\rangle\langle 1-f|. 
\label{Eq:HinF}
\end{align}
Conveniently, $H$ becomes block-off-diagonal in this representation. (For an odd-dimensional Hamiltonian, there will be an additional diagonal term of $|f=\frac{1}{2}\rangle\langle f=\frac{1}{2}|$ as discussed in Sec. \ref{Sec:Hig dim}). The Hamiltonian has an emergent `chiral' - like symmetry given by the unitary operator $\mathcal{Q}=\sigma_z$ (in even dimension), which anti-commutes with $H$. Here $\sigma_z$ is the Pauli-z matrix in the computation basis.

Consequently, the eigenvalues come in pairs as
\begin{align}
E_{\pm} &=\pm |E| e^{i\gamma}, 
\label{Eq:Heigenval}
\end{align}
where $|E|=\sqrt[4]{f(1-f)}\in \mathbb{R}^{+}$ (i.e., $f$-dependence is implicit in Eq. \eqref{Eq:Heigenval}). The two energy pairs are related to each other by $\gamma\rightarrow \gamma \pm\pi$, implying that they lie on diametrically opposite points on the complex energy plane. The energy amplitude $|E|$ is symmetric with respect to $f=\frac{1}{2}$. (For an odd-dimensional Hamiltonian $H$, an additional eigenvalue $\frac{1}{\sqrt{2}}e^{i\gamma_0}$ is present, whose absolute value is \textit{flat}, i.e., independent of the parameter $f$, see Sec. \ref{Sec:Hig dim}).

A comment is in order. Despite the block off-diagonal form of $H$ in the computational basis and the `chiral' \sout{`particle-hole'} pair of the energy eigenvalues, the approach does not lose its generality. It is true that an overall energy shift by the trace of $H$ (refer to Sec.~\ref{Subsec:General Example}) as $H\rightarrow H-{\rm Tr}(H)$, where ${\rm Tr}(H)\in\mathbb{C}$, leads to a violation of the anti-commutaion algebra in Eq. \eqref{Eq:h hdag one} since the operator $D$ is no longer proportional to the identity operator. However, it is obvious that ${\rm Tr}(H)$ gives a trivial overall (complex) shift to the entire energy eigenvalues while it does not contribute to the eigenstates. A modification to the above statement for a higher dimensional case is discussed in Sec. \ref{Sec:Hig dim}. 

The corresponding energy eigenstates can be expanded in the computational basis as
\begin{align}
\left|E_{\pm}\right\rangle &=\frac{1}{\sqrt{2|a|}}\left(|f\rangle \pm  a|1-f\rangle\right),~~{\rm where}~a=|a|e^{i\phi}.
\label{Eq:Heigenvec}
\end{align}
$a$ gives the complex expansion coefficient with phase $\phi$, and magnitude $|a| =\sqrt[4]{f/(1-f)}\in \mathbb{R}^{+}$. At each value of $f$, $|E_{\pm}\rangle$ are chiral partners of each other: $|E_{-}\rangle =\mathcal{Q}|E_{+}\rangle$. Note that the above eigenfunction is not yet normalized. For studying orthonormalization, we need to specify the dual space, which is discussed in Sec. \ref{Sec:Maps Dual Sym} below.  

Now, we interpret the two phases $\gamma$ and $\phi$. $\gamma$ is the winding angle in the complex energy phase; see Eq.~\eqref{Eq:Heigenval}. We see that the energy eigenstates arise from a linear superposition between corresponding states from the two halves of the interval, i.e., $\mathcal{F}_{0 \leqslant f \leqslant\frac{1}{2}}$ and $\mathcal{F}_{\frac{1}{2} \leqslant f \leqslant 1}$, where $\phi$ arises from the phase difference between these two halves, see Eq.~\eqref{Eq:Heigenvec}. Note that both $\gamma$ and $\phi$ are defined with respect to the exceptional point ($E=a=0$). 

For a given point in the computational space $F$, the description of the eigenvalues and eigenvectors of $H$ requires two complex parameters $E$, $a$. $|E|$, $|a|$ are determined by $f$, while $\gamma={\rm arg}(E)$, $\phi={\rm arg}(a)$ are two additional real parameters of $H$, which are dropped out in $F$. 

\subsection{Exceptional points and Normal point}\label{Subsec:EP NP}
There are three special points on the $0\le f\le 1$ domain where $H$ features interesting properties. $f=0,1$ corresponds to $|a|=0$, $\infty$, which are the exceptional points; whereas $f=\frac{1}{2}$, i.e., $|a|=1$ gives a normal point. (It is worth noting that what we call exceptional points are rather the $f=0,1$ \textit{surfaces} in the parameter space of the Hamiltonian.)  

An exceptional point is a spectral singularity
where two or more eigenvalues of a NH operator become identical, and the corresponding eigenstates coalesce into a single state. From Eq.~\eqref{Eq:Heigenvec}, we obtain exceptional points when the coefficient $|a|\rightarrow 0$ or $\infty$. They correspond to the \textit{coalescence} of the two energy eigenvectors $|E_{\pm}\rangle$ to a single state $|f\rangle$ or $|1-f\rangle$, respectively. This occurs at the two boundaries of the interval at $f=0,1$ where $|E|=0$. Notably, $f=0,1$ are the two contours of vacua in the computational basis (see Appendix~\ref{Appendix:EP as vacuum states}). 

Despite being perceived as a mathematical phenomenon until a few decades ago, exceptional points have found intriguing applications in a variety of physical phenomena including unidirectional light propagation~\cite{Yin2013, HuangShenMinFanVeronis}, high precision sensing~\cite{Wiersig:20}, level repulsion, quantum phase transition, quantum chaos, and others~\cite{Heiss_2012_Review, EPBerry, EPExp, EPLee, EPLaser, EPLaser2, EPHolographic, EPMagnon, EPExp2021}. Our observations provide a new perspective on the origin and tunability of the exceptional points. Specifically, our framework associates the origin of the exceptional points with the emergence of vacuum states, or equivalently the emergence of the fermion-like algebra (see Appendix~\ref{Appendix:EP as vacuum states}), in the computational basis. Furthermore, the exceptional points are tuned by the eigenvalue of the Hermitian operator, namely by the parameter $f$ or $|a|$ in our framework. More precisely, the distance of $f\in[0,1]$ from the \textit{normal point} at $f=1/2$, i.e., $|f-1/2|$, quantifies the strength of the non-Hermitian effects, or more accurately the ``non-normal effects"~\cite{NonNormalEffectsPhysRevResearch}, where the maximum of this distance corresponds to the occurrence of the exceptional points.

Finally, we define the normal point at $f=\frac{1}{2}$, giving $|a|=1$. Here, both $|f\rangle$ and $|1-f\rangle$ states are degenerate states of the $F$ operator. The commutator is $[H,H^{\dagger}]=I-2F$, and hence $H$ and $H^{\dagger}$ commute at this point. We call this point a {\it normal point}. Moreover, at the normal point, $|a|=1$, $|E_{\pm}\rangle$ achieve maximal quantum coherence in the computational basis.  The details of the degenerate case are given in Appendix \ref{Appendix:NP in 2D}.

\section{Maps, Dual Space, and Symmetries}\label{Sec:Maps Dual Sym}
In NH quantum theory, the usual connection between eigenstates and their duals via Hermitian conjugation breaks down. The popular biorthogonal eigenstate is obtained by solving the eigenstates of $H^{\dagger}$, and yet there is no simple map between the eigenstates of $H$ and $H^{\dagger}$. Special cases arise for $\mathcal{PT}$-symmetric or pseudo-Hermitian Hamiltonians, which do not extend to their symmetry-broken regions. We emphasize that our dual space map $-$ defined by a discrete transformation in the computational space $-$ is generic and works in the $\mathcal{PT}$- symmetric and pseudo-Hermiticity broken and unbroken regions. We also find weaker constraints on real energy eigenvalues.

\subsection{Dual space}\label{Subsec:Dual}
\begin{figure}[t]
  \centering
  \includegraphics[width=7cm]{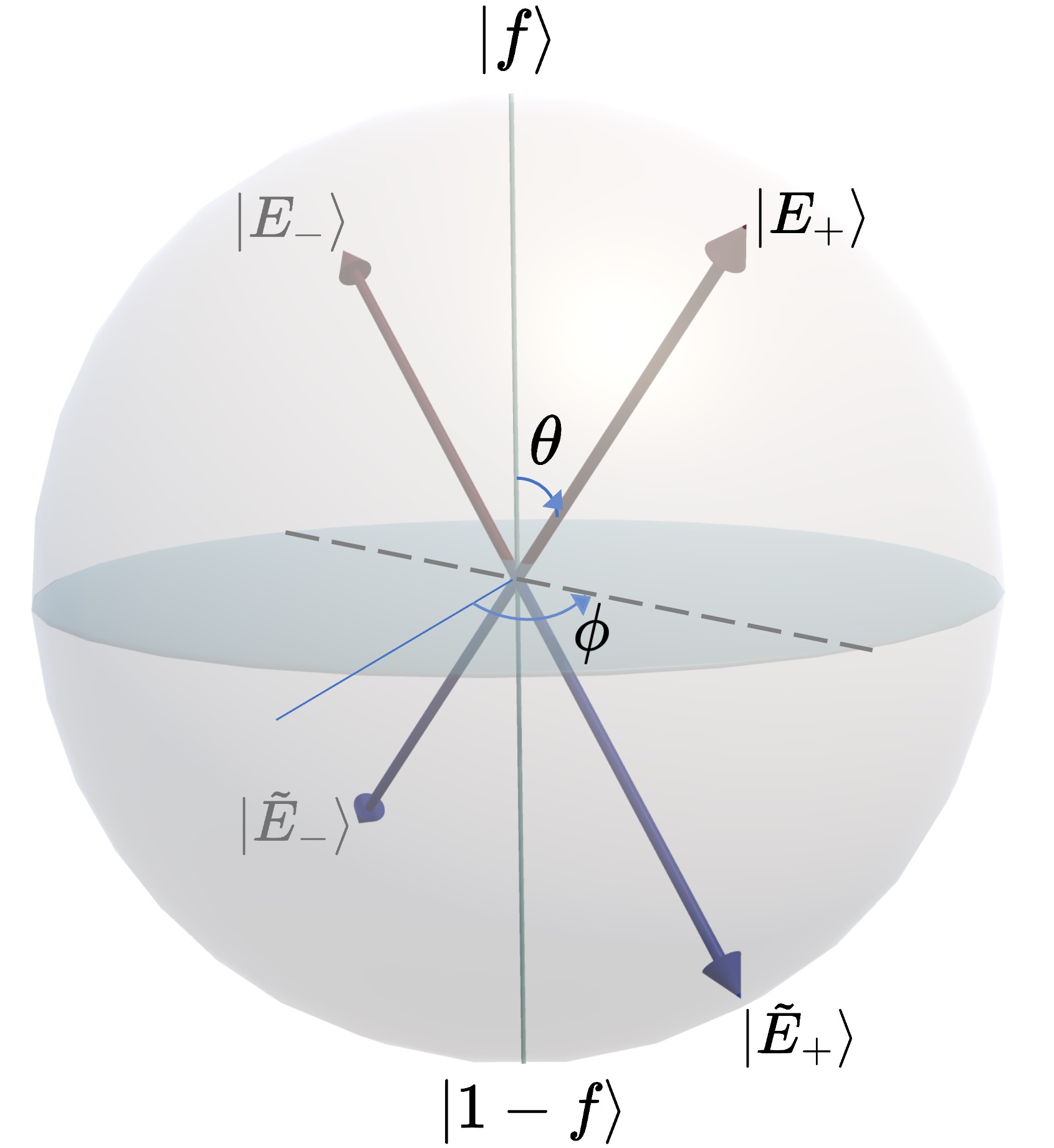}
  \caption{Energy eigenstates and the dual states are exhibited on the computational Bloch sphere parameterized by expressing the expansion parameter $a={\rm tan}\frac{\theta}{2} e^{i\phi}$. Here the exceptional points/contours are located at the two poles, while the normal point (contour) lies at the equator. The two energy eigenstates $|E_{\pm}\rangle$ lie in the same hemisphere, demonstrating that they are not orthogonal to each other, except at the equator (normal point), while collapsing to a single basis state at the poles (exceptional points/contours). The biorthogonal dual states $|\tilde{E}_{\pm}\rangle$ lie at the antipodal points to $|E_{\mp}\rangle$ in this computation space. }
  \label{Fig:Bloch sphere}
\end{figure}

Before plunging into redefining the dual space of $|E_{\pm}\rangle$, we remind the readers why the traditional Hermitian conjugation map fails here. It is convenient to represent the dimensionless parameter $|a|$ by an angular variable $\theta$ as 
\begin{equation}
|a|={\rm tan}\frac{\theta}{2},
\label{eq:aintheta}
\end{equation}
where $0\le \theta\le \pi$. This unveils a geometric view of $|E_{\pm}\rangle$, living on a hypersphere $\mathbb{S}^2$ (for a given value of $f$) parameterized by $\phi$ and $\theta$, as shown in Fig. \ref{Fig:Bloch sphere}. The computational basis $|f\rangle$ and $|1-f\rangle$ lie along the two antipodal points, as expected by virtue of their orthogonality. However, $|E_{\pm}\rangle$ states do \textit{not} align diagonally opposite to each other and hence do \textit{not}, in general, form an orthogonal basis. The two states are connected by $\phi\rightarrow \phi+\pi$ with fixed $\theta$. They become orthogonal to each other only at $\theta=\pi/2$, which is the \textit{normal point} ($|a|=1$) (see more discussions in the Supplemental Material~\cite{Supp}). At $\theta=0$ or $\theta=\pi$ (i.e., $|a|=0$ or $|a|=\infty$ at $f=0$ and 1), $|E_{\pm}\rangle$ collapse to either $|f\rangle$ or $|1-f\rangle$ states, respectively $-$ which are the two \textit{exceptional points}.

Now our job is well defined. We need to find two states $|\tilde{E}_{\pm}\rangle$, which lie at the two corresponding antipodal points of $|E_{\mp}\rangle$. The corresponding states are
\begin{equation}
|\tilde{E}_{\pm}\rangle=\sqrt{\frac{|a|}{2}}\left(|f\rangle \pm \frac{1}{|a|} e^{i\phi}|1-f\rangle\right),
\label{Eq:biortho}
\end{equation}
It is easy to check that $|\tilde{E}_{\pm}\rangle$ are the eigenstates of $H^{\dagger}$ with eigenvalues $\pm E^*$. They are bi-orthonormal states following $\langle \tilde{E}_{n}|E_{m}\rangle=\delta_{nm}$, where $n,m=\pm$, and $\langle \tilde{E}_{n}|$ corresponds to the usual adjoint operation. 

Can we relate the two antipodal points by any unitary and/or antiunitary transformation? We prove below that there exist dynamical discrete `space-time' transformations in the computational space. $|E_\pm\rangle$ are related to $|\tilde{E}_\pm\rangle$ by a \textit{reflection} about the $\theta =\frac{\pi}{2}$ (i.e., $|a|=1$) or `normal' plane: $\theta\rightarrow \pi-\theta$ (i.e., $|a|\rightarrow |a|^{-1}$), $\phi\rightarrow\phi$. On the other hand, $|E_\pm\rangle$ are related to $|\tilde{E}_\mp\rangle$ by an \textit{inversion} operation: $\theta\rightarrow \pi-\theta$, and $\phi\rightarrow \phi+\pi$. We define the corresponding two operations by unitary and anti-unitary operators as: (i) $|\tilde{E}_\pm\rangle  \propto  \mathcal{U}_1 |E_\pm\rangle$, (ii) $|\tilde{E}_\pm\rangle  \propto  \mathcal{U}_2 |E_\mp\rangle$, (iii) $|\tilde{E}_\pm\rangle  \propto  \mathcal{A}_1 |E_\pm\rangle$, and (iv) $|\tilde{E}_\pm\rangle  \propto  \mathcal{A}_2 |E_\mp\rangle$, where $\mathcal{U}_i$ and $\mathcal{A}_i =\mathcal{V}_i\mathcal{K}$ with $\mathcal{K}$ is the complex conjugation, are unitary and anti-unitary operators, respectively. 

    \subsubsection{\textit{Reflection operation ($\mathcal{U}_1$):}}
    The unitary operator $\mathcal{U}_1$ corresponding to $\mathcal{U}_1 |E_\pm\rangle := \pm |\tilde{E}_\pm\rangle$ is defined in the computational space as
\begin{align}\label{Eq:U1 in F}
          \mathcal{U}_1=e^{i\phi} |1-f\rangle \langle f| +  e^{-i\phi} |f\rangle \langle 1-f|.
\end{align}
where $\phi$ is the same phase as in Eq.~\eqref{Eq:Heigenvec}. Its matrix-valued expression is $\mathcal{U}_1=\cos{\phi}\sigma_x + \sin{\phi}\sigma_y$. We make two important observations here. Firstly, $\mathcal{U}_1^2=\mathbb{I}$, implying $\mathcal{U}_1$ is a {\it discrete} transformation. Secondly, 
\begin{align}\label{Eq:pm norm U1}
    \langle \mathcal{U}_1 E_m |E_n\rangle  &=(-1)^n\delta_{mn},
\end{align}
where $n=1,2$ correspond to $|E_{\mp}\rangle$ states, respectively. The negative norm indicates that the energy states live on a non-Riemannian manifold. We define a Hamiltonian-specific metric, which is equivalent to defining a Hamiltonian-specific operator with the desired property that $\mathcal{C}_1 |E_n\rangle := (-1)^n |E_n \rangle$. This makes the length of the eigenvectors unity:
\begin{align}\label{Eq:Pos norm U1C1}
\langle \mathcal{C}_1^{\dagger} \mathcal{U}_1 E_n| E_m\rangle  = \delta_{mn}.
\end{align}
This operator expressed in the computation basis is 
\begin{eqnarray}
    \mathcal{C}_1 
    &=& a|1-f\rangle \langle f| + a^{-1}|f\rangle \langle 1-f|.
    \label{Eq:C1 in F}
\end{eqnarray}
By construction $[H,\mathcal{C}_1]=0$, i.e. $\mathcal{C}_1$ is a \textit{``hidden-symmetry"} of the system, in analogy with the \PT- quantum theory literature \cite{BenderReview}.

Therefore, we deduce that $\mathcal{C}_1^{\dagger}\mathcal{U}_1$ is a non-unitary dynamical {\it metric} operator obtained to be 
\begin{equation}
\mathcal{C}_1^{\dagger}\mathcal{U}_1=|a||f\rangle \langle f| + |a|^{-1} |1-f\rangle \langle 1-f|.
\end{equation}
The physical interpretations of these two operators are evident. $\mathcal{C}_1^{\dagger}\mathcal{U}_1$ is like a metric that adjusts the `length' of the two basis states, where $\mathcal{C}_1$ incorporates a `probability flow' (decoherence), and $\mathcal{U}_1$ incorporates the geometric `phase obstruction' in the computational space.  



    \subsubsection{\textit{Inversion operation ($\mathcal{U}_2$):}}
        Proceeding similarly, we obtain $\mathcal{U}_2=-i(\sin{\phi}\sigma_x - \cos{\phi}\sigma_y)$ with $\mathcal{U}_2|E_\pm\rangle=\pm|\tilde{E}_\mp\rangle$ and $\mathcal{U}^2=-\mathbb{I}$. $\mathcal{U}_2$ and $\mathcal{U}_1$ are related by the `particle-hole' symmetry: $\mathcal{U}_2=\mathcal{Q}\mathcal{U}_1$. A similar negative norm problem is remedied by defining a $\mathcal{C}_2$ operator, which turns out to be $\mathcal{C}_2=-\mathcal{C}_1$. In what follows $\langle \mathcal{C}_2^{\dagger} U_2 E_\mp |$ gives the \textit{same} dual space corresponding to $|E_\pm\rangle$. We observe that at the normal point ($|a|=1$), $\mathcal{C}_1^{\dagger}\mathcal{U}_1=\mathbb{I}$ and $\mathcal{C}_2^{\dagger}\mathcal{U}_2=\mathcal{Q}$. Then the above inner product definitions coincide with the usual Hermitian conjugation dual space.

    \subsubsection{\textit{Anti-unitary operation ($\mathcal{A}_i$):}}
    Similar to the case of unitary operators, we can define anti-linear operators $\mathcal{A}_1=\sigma_x \mathcal{K}$ with $\mathcal{A}_1 |E_\pm\rangle = \pm e^{-i\phi} |\tilde{E}_\pm\rangle$, and $\mathcal{A}_2=i\sigma_y \mathcal{K}$ with $\mathcal{A}_2 |E_\pm\rangle = \pm e^{-i\phi} |\tilde{E}_\mp\rangle$. They are related to each other by $\mathcal{A}_2=\mathcal{Q}\mathcal{A}_1$. We observe that $\mathcal{A}_i^2=(-1)^{i+1}\mathbb{I}$. For $\mathcal{A}_1$ and $\mathcal{A}_2$, we find that $\langle \mathcal{C}_3^\dagger \mathcal{A}_1 E_\pm|$ and $\langle \mathcal{C}_4^\dagger \mathcal{A}_2 E_\mp|$ are the corresponding duals of $|E_\pm\rangle$ with $\mathcal{C}_3=e^{-i\phi}\mathcal{C}_1$ and $\mathcal{C}_4=-\mathcal{C}_3$.


\subsection{Symmetry-enforced real energy}\label{Subsec:Sym}
How do $H$ and $H^{\dagger}$ transform under these maps, and do these transformations manifest any symmetry of the system? Interestingly, $\mathcal{U}_i$ and $\mathcal{A}_i$ provide a relation between $H$ and $H^{\dagger}$ as
\begin{eqnarray}
\label{Eq:UHHdag}
\mathcal{U}_i H \mathcal{U}_i^{-1}&=&(-1)^{i+1}e^{2\rm{i}\gamma}H^\dagger,\\
\mathcal{A}_i H \mathcal{A}_i^{-1}&=&(-1)^{i+1} H^\dagger,
\label{Eq:AHHdag}
\end{eqnarray}
where $i=1,2$. These expressions resemble the pseudo-Hermitian criterion~\cite{MostafazadehReview, MostafazadehReview1a, MostafazadehReview1b, MostafazadehReview2}, but are generalized here for any random Hamiltonian. This gives a clue that a combination of $\mathcal{U}_i$ and $\mathcal{A}_i$ can perhaps be made to be a symmetry of the system. Among the four possible combinations, we have two independent combinations: $\mathcal{S}_1= \mathcal{A}_1^{-1}\mathcal{U}_1 = \mathcal{A}_2^{-1}\mathcal{U}_2=(\cos\phi \mathbb{I}-i\sin\phi\sigma_z)\mathcal{K}$ and $\mathcal{S}_2=\mathcal{A}_1^{-1}\mathcal{U}_2 = \mathcal{A}_2^{-1}\mathcal{U}_1=i(\sin\phi \mathbb{I}+i\cos\phi\sigma_z)\mathcal{K}$, where $\mathcal{S}_1^2=\mathcal{S}_2^2=I$. $\mathcal{S}_i$ are diagonal in the computational basis and transform the Hamiltonian as 
\begin{eqnarray}
\mathcal{S}_i H \mathcal{S}_i^{-1}=(-1)^{i+1} e^{-2\rm{i}\gamma}H,
\label{Eq:SHH}
\end{eqnarray}
for $i=1,2$. Interestingly, for either \textit{real} or \textit{imaginary} energies, one of the $\mathcal{S}_i$s becomes a \textit{symmetry}, and the other becomes an \textit{anti-symmetry}.

\begin{table}[t] 
\caption{\label{Table:Symmetry table}We classify the NH Hamiltonians according to the generalized Bernard LeClair (gBLC) classification scheme. The values $\pm 1$ correspond to $\epsilon_{\Xi}=\pm$ in Eq.~\eqref{Eq:GBL}, while $0$ means either the transformation operator does not exist or is a dynamical operator as in Eq.~\eqref{Eq:UHHdag}. The $\mathcal{Q}$ and $\mathcal{A}_1$ symmetries are always present as they are static symmetries in the computational basis (see main text), while the $\mathcal{U}_1$ symmetry class varies according to whether the energy eigenvalues are real, imaginary, or complex, i.e. on the values of $\gamma$ in Eq.~\eqref{Eq:UHHdag}. The same results for $\mathcal{A}_2$, $\mathcal{U}_2$ and $\mathcal{S}_2$ are obtained by the relations $\mathcal{A}_2=\mathcal{Q}\mathcal{A}_1$, $\mathcal{U}_2=\mathcal{Q}\mathcal{U}_1$, and $\mathcal{S}_2=-\mathcal{Q}\mathcal{S}_1$.}
\begin{tabular}{|c|c|c|c|c|}
\hline
Energy  & $\mathcal{Q}$
& $\mathcal{A}_1$ & $\mathcal{U}_1$ & $\mathcal{S}_1$ \\ 
 & (P-class) & (C-class) & (Q-class) & (K-class)\\

\hline
Complex            & -1                           & +1                               & 0        & 0                       \\
Real               & -1                           & +1                               & +1  &+1                            \\
Imaginary          & -1                           & +1                              & -1      & -1                         \\ \hline
\end{tabular}
\end{table}

For \textit{real} energies, which correspond to $\gamma=n\pi$ for $n\in \mathbb{Z}$,  $\mathcal{S}_{1}$ ($\mathcal{S}_{2}$) is a symmetry (anti-symmetry) of $H$. For  \textit{imaginary} energies, i.e., 
$\gamma=\frac{n\pi}{2}$ with $n$ to be an odd integer, $\mathcal{S}_1$ ($\mathcal{S}_2$) becomes an anti-symmetry (symmetry) of $H$.  For other values of $\gamma$, i.e., for general complex energies, $\mathcal{S}_i$ are not symmetries/antisymmetries of $H$, yet $\mathcal{U}_i$ and $\mathcal{A}_i$ provide linear and antilinear maps to the dual space, respectively. 

Eqs.~\eqref{Eq:UHHdag}, \eqref{Eq:AHHdag}, and \eqref{Eq:SHH} can be used to classify the Hamiltonian in the generalized Bernard LeClair (gBLC) classification scheme~\cite{BLC,BdG2NH,NHClassification1,NHClassification2,NHClassification3,VishwanathNH}. For a unitary symmetry operation $\Xi$, if the Hamiltonian transforms as 
\begin{equation}
H =\epsilon_{\Xi}\Xi f(H)\Xi^{-1},  
\label{Eq:GBL}
\end{equation}
and if $\epsilon_{\Xi}=\pm 1$ and $f(H)$ equal to $H$, $H^T$, $H^\dagger$ and $H^*$, then such symmetry classes are named as P, C, Q, and K classes, respectively (for the P-class $\epsilon_{\Xi}=+1$ is a trivial unitary transformation). According to our definition, the `particle-hole' symmetry $\mathcal{Q}$ corresponds to the P-class, while $\mathcal{V}_i$, $\mathcal{U}_i$ and $\mathcal{S}_i$ belong to the C-, Q- and K-classes, respectively. We list the symmetries of our Hamiltonian in Table \ref{Table:Symmetry table}. $\mathcal{Q}$ is, by construction, a static symmetry in the computational basis. $\mathcal{A}_i$ are also static operators in the computational basis and emerge due to the constraint in Eq.~\eqref{Eq:h hdag one}. $\mathcal{U}_i$ is only present as a gBLC symmetry operation if the energy eigenvalues are real or imaginary. Since $S_i$ is a product of $\mathcal{U}_i$, and $\mathcal{A}_i$, they form a group structure. 

$\mathcal{S}_i$ operators are defined on the computational basis, and their forms on the original Hamiltonian basis are Hamiltonian-specific. Therefore, $\mathcal{S}_1$ is a general symmetry, constraining real eigenvalues, which may or may not coincide with the $\mathcal{PT}-$ symmetry operation\cite{BenderReview}.  Similarly, the dynamical metric operators $\mathcal{C}_i^{\dagger}\mathcal{U}_i$ behave analogous to the pseudo-Hermitian class but are not the same and work in the pseudo-Hermiticity broken regions also\cite{MostafazadehReview, MostafazadehReview1a, MostafazadehReview1b, MostafazadehReview2}. We emphasize that $\mathcal{S}_{1,2}$ are `dynamical' operators as they depend on the Hamiltonian parameter $\phi$, and their realization as symmetry/anti-symmetry depends on the parameter values $\phi$ and $\gamma$. 

However, interesting special cases of `static' global symmetries can be worked out that rule out the possibility of complex energies constraining them to be purely real/imaginary. For instance, the operator $\mathcal{S}=\sigma_z\mathcal{K}$ as a global symmetry constraints $\gamma$ and $\phi$ as $\gamma=n\pi/2$ and $\phi=\gamma+(m+1/2)\pi$ where $n,m\in\mathbb{Z}$. If this operator remains static in the original Hamiltonian basis, it coincides with the $\mathcal{PT}$-operator with our formalism remaining valid in the $\mathcal{PT}$-broken regime as well. On the other hand, if this operator becomes dynamical in the original basis, this is a more general global symmetry than the $\mathcal{PT}$-operation which is still static in the computational basis.


\section{Parametrization, Applications, and Examples}\label{Sec:App Example}
We discuss how the computational space, energy eigenspaces, dual space, and complex energy space are embedded in the parameter space of random NH Hamiltonians. In this process, we also demonstrate the efficacy of our framework in distilling the role of each Hamiltonian parameter in the physical properties of general interests. We subsequently consider several representative examples to demonstrate our formalism. 
 
\subsection{General NH Hamiltonian and roles of its parameters}\label{Subsec:General Example}
We start with a general $N\times N$ random Hamiltonian $H$ which has $N^2$ complex variables $h_i$. The trace of a Hamiltonian and the global scaling of the eigenvalues ($|h|$) do not contribute to the eigenstates. So, we consider the traceless part of the Hamiltonian with $N^2-1$ complex parameters. What can be a general operator with eigenvalue $\propto |h|$ which commutes with $H$? Our claim in Eq.\eqref{Eq:h hdag one} is that the operator $\{H,H^{\dagger}\}$ does this job for a wider class of random Hamiltonians. This is certainly true for two-dimensional Hamiltonians as we explore first.

We consider a 2D traceless random NH Hamiltonian expanded on the basis of Pauli matrices with complex coefficients as
\begin{equation}\label{Eq:H dot sigma}
H=\Vec{h} \cdot \Vec{\sigma},~{\rm where}~ \Vec{h}=\left(h_x,h_y,h_z\right)\in \mathbb{C}.
\end{equation}
The Hamiltonian follows $\left\{H,H^\dagger\right\}=2|\Vec{h}|^2\mathbb{I}$. The scaling parameter $d=2|\Vec{h}|^2$ is set to 1 for simplicity. 

\begin{figure}[t]
    \centering
    \includegraphics[width=7cm]{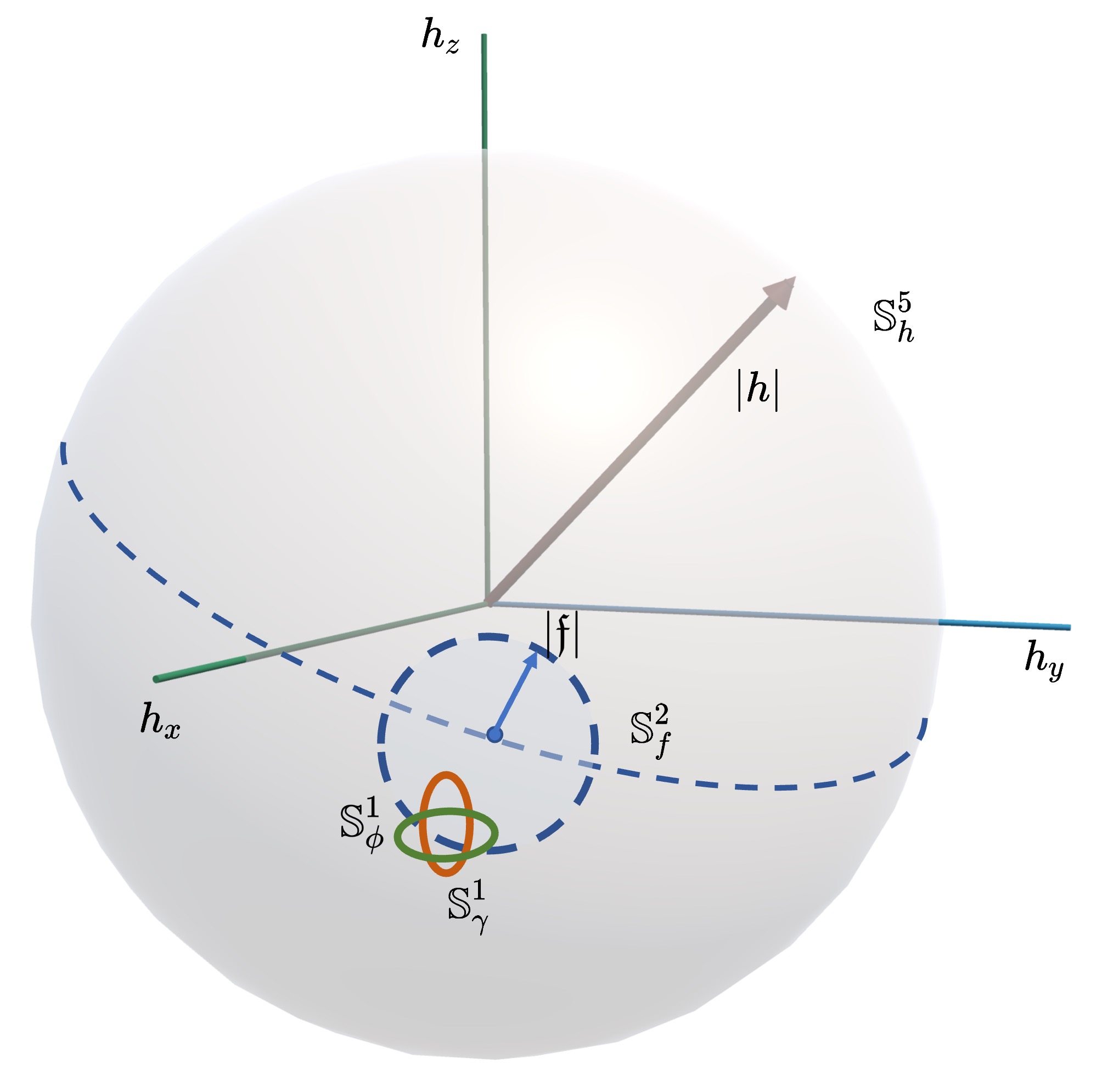}
    \caption{Schematic representation of the splitting of the complex parameter space of the Hamiltonian into the computational Hilbert space $\mathbb{S}_\mathfrak{f}^{2}$, expansion parameter space $\mathbb{R}'_{|\mathfrak{f}|}$, $\mathbb{S}_{\phi}^1$, and the complex energy space $\mathbb{S}^1_{\gamma}$. The radius of the hypersphere $|h|$ corresponds to the scaling parameter $d=2|h|^2$. The blue thin dashed hypercycle is the $|\mathfrak{f}|=0$ contour of the normal point. The computational space "\textit{encircles}" the normal point as shown by the thick blue dashed line and extends up to $|\mathfrak{f}|=1/2$ contour. $\gamma$ and $\phi$ are defined with respect to the exceptional point/contour. }
    \label{Fig:H param space}
\end{figure}

The computational space spans a subset of the Hamiltonian's parameter space, parameterized by $N^2-1=3$ real parameters. This can be easily deduced by expressing the Hermitian $F$ operator as $F= \frac{\mathbb{I}}{2} + \Vec{\mathfrak{f}}\cdot \Vec{\sigma}$,  where $\Vec{\mathfrak{f}}\in \mathbb{R}_{\mathfrak{f}}^3$ and is given by $\mathfrak{f}_{\mu}=-2\epsilon_{\mu\nu\rho}{\rm Im}[h^*_{\nu}h_{\rho}]$ with $\mu,\nu,\rho=x,y,z$ (no summation convention is implied on the repeated indices here). The eigenvalues of $F$ are determined by $|\mathfrak{f}|\in \mathbb{R}_{|\mathfrak{f}|}$ as  $f=\frac{1}{2}+|\mathfrak{f}|$, and $1-f=\frac{1}{2}-|\mathfrak{f}|$. However, there is a restriction on the allowed values of $|\mathfrak{f}|\in \mathbb{R}^{\prime}_{|\mathfrak{f}|}$, where the prime symbol denotes its limit $0\le |\mathfrak{f}|\le \frac{1}{2}$. The basis vectors $|f\rangle$, $|1-f\rangle$ lie at the antipodal points of $\mathbb{S}_\mathfrak{f}^{2}$, parameterized by the two angular variables on the $|\mathfrak{f}|$ surface.  


The Hamiltonian's eigenvalues $E$ depends on $|\mathfrak{f}|$, but the phase $\gamma={\rm arg}(E)\in \{\Vec{h}\}-\{\Vec{\mathfrak{f}}\}$. The energy states' expansion parameter in the computation basis is $a$, with $|a|$ is determined by $|\mathfrak{f}|$, while its phase $\phi={\rm arg}(a)\in \{\Vec{h}\}-\{\Vec{\mathfrak{f}}\}$.

 
The dual space is defined by the transformation in the Hamiltonian parameters as 
$|a|\rightarrow|a|^{-1}$, $\phi\rightarrow \bar{\phi}$ where $\bar{\phi}=\phi~(\phi+\pi)$ for $\mathcal{U}_1$ ($\mathcal{U}_2$) symmetries (and the same for $\mathcal{A}_1$ ($\mathcal{A}_2$)). The operator $\mathcal{C}_i$ in the inner product depends on $a$, i.e., $\theta$ and $\phi$ in the computational basis.




\subsection{Examples and Insights}\label{Subsec:Example alpha beta}
\subsubsection{Gain/Loss Hamiltonians, $\mathcal{PT}$ symmetry and beyond}
A quintessential two-level system with gain and loss, which encompasses several physical systems 
is of the following general form \cite{PhysRevA.98.042117}:
\begin{equation}
H=\begin{pmatrix}
-i \gamma_1 & \Omega_r-i \Omega_i \\
\Omega_r+i \Omega_i & \Delta-i \gamma_2
\end{pmatrix},
\label{eq:exmH1}
\end{equation}
where $\Delta,\gamma_{1,2},\Omega_{r,i} \in \mathbb{R}$. We set $\hbar=1$. $\Delta$ is the detuning parameter or the onsite energy difference, while $\Omega=\Omega_{r}+i\Omega_i$ constitutes the complex tunneling between the two levels. $\Delta$ and $\Omega$ are Hermitian parameters intrinsic to the system. 
The NH parameters $\gamma_{1,2}$ control the gain/loss terms, among which only their difference $\delta\gamma=(\gamma_1-\gamma_2)/2$ enters into $|\mathfrak{f}|$ (i.e., $f$, $a$ and $E$). $(\gamma_1+\gamma_2)/2$, which enters into ${\rm Tr}(H)$, gives a (global) complex energy shift to all eigenenergies, and hence we shift the Hamiltonian to $H\rightarrow H-{\rm Tr}(H)$. We set $\Delta=0$ for simplicity, and it can be inserted back into the final result with physical intuition. So, we have one Hermitian parameter $\Omega$ and one NH parameter $i\delta\lambda$. 

The global energy scaling parameter $d=2(|\Omega|^2+|\delta \lambda|^2)$ is set to 1, which is equivalent to scaling $\Omega\rightarrow\Omega/\sqrt{d}$, and $\delta\gamma\rightarrow \delta\gamma/\sqrt{d}$. 
From the analysis of Sec.~\ref{Subsec:General Example}, we easily identify $\mathfrak{f}_{x,y}=\pm2(\delta\lambda)\Omega_{i,r}$, $\mathfrak{f}_z=0$. The system's intrinsic Hermitian parameter $\Omega$, rather ${\rm arg}(\Omega)$,  determines the computational states, while the NH parameter $\delta\gamma$ (globally) scales its eigenvalue $|\mathfrak{f}|$. In fact, the $F$  operator is a Hermitian (massive) SSH-like model with eigenvalues $f=\frac{1}{2}\pm 2|\delta\gamma||\Omega|$ and a Berry phase related to ${\rm arg}(\Omega)$. $\delta\gamma$ appears as a gap between their eigenvalues, giving a topological transition point at $\delta\gamma=0$, where $H$ becomes a normal operator (or Hermitian if we ignore its trace). The exceptional points occur by tunning to $4|\delta\gamma||\Omega|=1$. 
Energy eigenvalues are $|E|=\sqrt[4]{1/4-4|\delta\gamma|^2|\Omega|^2}$. The phase of the energy states changes between $\phi=(2n+1)\pi/2$ and $n\pi$ for $n\in\mathbb{Z}$ as we cross the exceptional point. 

We notice that both energies are either real or imaginary in different parameter regimes. Is there any symmetry (\PT~ or pseudo-Hertimicity, etc) that protects these regions?  Our dual space map $\mathcal{S}_i$ easily gives the answer. From the definition, we find $\mathcal{S}_1=\sigma_z\mathcal{K}$, and $\mathcal{S}_2=\mathcal{K}$ for real eigenvalues. $\mathcal{S}_1$ coincides with the \PT~ symmetry operator in the computational basis. Importantly, \PT~ symmetry is broken at the exceptional point, and the \PT~theory does not work in the \PT-broken region. However, our formalism works in both the \PT~ broken and unbroken regions. 

The Hamiltonian in Eq.~\eqref{eq:exmH1} is a prototypical one in the laser, coherent photo-absorbers,\cite{antilaser1} or their \PT-symmetric combinations,\cite{ptlaser1,ptlaser2} in photonics, quantum optics,\cite{qoptics}  qubits,\cite{qubit} or open quantum dots (zero-spatial dimension) studies.\cite{qdots,UedaReview} In one-spatial-dimension, Eq.~\eqref{eq:exmH1} coincides with the NH SSH model for $\Omega=u+ve^{ik}$, where $k$ is crystal momentum and $u$, $v\in\mathbb{R}$ are nearest neighbor tunneling matrix elements.\cite{GhatakReview,UedaReview,SSHLee,SSH2} In two-spatial-dimension, it represents the NH Chern insulator, and/or Rashba-spin-orbit coupling, or the $p+ip$ superconducting Hamiltonians\cite{GhatakReview,UedaReview,GhatakSC,NHChern1,NHChern2} for $\Omega=\sin{k_y}+i\sin{k_x}$, and $\Delta=2(m+\cos{k_x}+\cos{k_y}$), where $m\in \mathbb{R}$. 

\subsubsection{Non-reciprocal hopping}
Our second example is also a popular one discussed in many branches of quantum physics, where the coupling between the two levels is non-reciprocal. Omitting any gain/loss term here, the prototypical non-reciprocal Hamiltonian is  
\begin{equation}
    H=\begin{pmatrix}
0 & \Omega_1 \\
\Omega_2 & \Delta
\end{pmatrix},
\label{eq:exmH2}
\end{equation}
where $\Delta\in \mathbb{R}$. First, we consider $\Omega_{1,2}\in\mathbb{R}$, in which case, with a unitary rotation by $e^{-i\pi\sigma_x/4}$, we find this Hamiltonian coincides with the Hamiltonian in Eq.~\eqref{eq:exmH1} by identifying $\Omega_r= (\Omega_1+\Omega_2)/2$, $\Omega_i=\Delta/2$, and $\delta\gamma=(\Omega_1-\Omega_2)/4$. The key message is that the only NH parameter of this model is the degree of non-reciprocity, i.e., anisotropic hopping between the two levels. The computational basis,  distance from the exceptional point, topology, dual space maps, and symmetries belong to the same classification scheme as the Hamiltonian with gain and loss terms.  Such a model appears in unidirectional/nonreciprocal transition/reflection or currents \cite{nonrechop,nonrechop2} open circuits,\cite{circuit} scattering matrices,\cite{NHQM} and generic time-reversal broken transport properties\cite{transport1,transport2}

In one or higher spatial dimensions, classes of such Hamiltonian coincide with a bipartite Hatano-Nelson model,\cite{HatanoNelson} non-reciprocal superconductivity,\cite{GhatakSC} and models featuring skin effects.\cite{skineffect1,skineffect2} The generic form of the tunnelings in one-spatial dimension are $\Omega_1=t_R+t_Le^{-ik}$, $\Omega_2=t_L+t_Re^{ik}$, both being complex here. By going to the computational space: $\mathfrak{f}_x=-\delta t\Delta (1-\cos k)$, $\mathfrak{f}_y=\delta t\Delta\sin k$, $\mathfrak{f}_z=0$: we discover that there is only one NH parameter of importance in this theory, which is $\delta t=(t_R-t_L)/2$. Since $\delta t\Delta$ poses a global scaling to the computational space, its basis depends on the phase winding in the momentum space $k$, similar to the Hermitian SSH or Chern insulator Hamiltonian. Mathematically, the role of $\delta t$ turns out to be equivalent to $\delta\gamma$. 

The above examples demonstrate the distillation of essential NH parameters that dictate the physical properties associated with non-Hermiticity in open quantum systems. More examples, including a general two NH parameter Hamiltonian and application of our formalism in Hermitian Hamiltonians, are discussed in Appendix~\ref{Appendix:MoreExamples}.

\begin{figure*}[ht]
    \includegraphics[width=14cm]{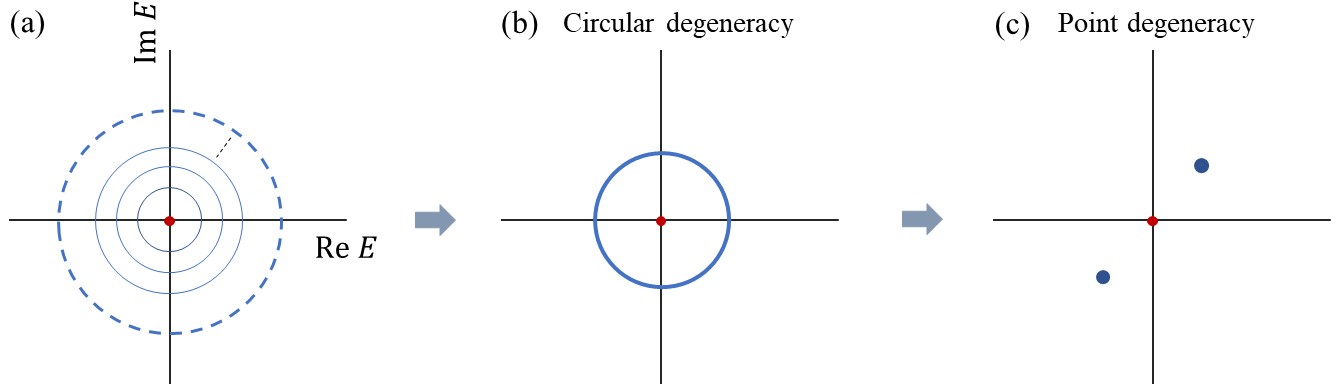}
    \caption{(a) Energy spectra in a higher dimensional Hamiltonian are schematically plotted as concentric circles in the complex energy plane for the case where energy eigenvalues do not cross. Each circle  corresponds to a different $f_n$ with the eigenvalue pair $\pm |E_n|e^{i\gamma_n}$ lying on their diametrically opposite points. The dashed blue circle gives the maximum radius $|E_n|=1/\sqrt{2}$ corresponding to the normal point, where the center of the complex plane is the exceptional point. (b) \textit{Circular degeneracy:} In this case, two or more energy levels coincide on the same circle $|E|$, but differ by the phases $\gamma_n$. (c) \textit{Point degeneracy:} This occurs when two or more energy levels possess the same amplitude and phase, i.e., they correspond to a single point on the complex plane (with the `particle-hole' component lying at the diametrically opposite point). }
    \label{Fig:Spectral Flattening}
\end{figure*}

\section{Higher dimension and degeneracy}\label{Sec:Hig dim}

In the two-dimensional case, the `particle-hole' eigenvalue pair $\pm |E|e^{i\gamma}$ is scaled by a single parameter $d$ and the corresponding computational space is parameterized by $f$. Generalization to any higher dimension requires considerations of the classifications of the eigenspectrum of $D$ and $F$ operators and will be pursued in future studies. Here we focus on the $P$-class of higher dimensional Hamiltonians with `particle-hole' symmetry (see Table \ref{Table:Symmetry table}) in which all eigenvalues are separated from each other in the complex energy space (see Fig.~\ref{Fig:Spectral Flattening}(a)). This means there exists a block diagonal form of the Hamiltonian for each eigenvalue pair $\pm |E_n|e^{i\gamma_n}$ with the corresponding scaling and computational parameters $d_n$ and $f_n$, respectively. The full basis states are a collection of all the computational basis states, and the expansion coefficients are $a_n=|a_n|e^{i\phi_n}$, where the definitions of $|E_n|$ and $|a_n|$ remain the same in terms of $f_n$.  

Interesting special cases arise for two degenerate cases, apart from the exceptional points. (1) An $m$-fold degeneracy of the eigenvalue pairs $\pm |E|e^{i\gamma_n}$, $n=1$ to $m$, where the phases $\gamma_n$ are different but amplitude $|E|$ is the same. We call it the \textit{circular degeneracy} in the complex energy plane, see Fig.~\ref{Fig:Spectral Flattening}(b). (2) A second simpler case is when all the phases are also the same to $\gamma$, which we term as the \textit{point degeneracy} as shown in Fig.~\ref{Fig:Spectral Flattening}(c). 

\begin{enumerate}
\item \textit{Circular degeneracy:} Since $|E|$ only depends on $f$ (and $d$, which we set to be 1 as before), the computational basis $|f\rangle$ is $m$-fold degenerate. We choose an $m$-fold degenerate orthogonal basis $|f\rangle_i$ and $|1-f\rangle_i$, and express the corresponding Hamiltonian in the computational manifold as block off-diagonal form   
\begin{eqnarray}\label{Eq:HinF hig dim deg f}
        H=\sum_{i,j=1}^m A_{ij} |1-f\rangle_i \langle f|_j + B_{ij} |f\rangle_i \langle 1-f|_j,
 \end{eqnarray}
where $A_{ij}$, $B_{ij}\in \mathbb{C}$. Comparing this to Eq.~\eqref{Eq:HinF} it is obvious that the eigenvalues of $A$ and $B$ give the factors in the ladder operations in Eq.~\eqref{Eq:ladder action d}.  
Expressing $|E^n_{\pm}\rangle$ in a spinor field $(\psi^n_A~\pm\psi^n_B)^T$ we obtain $A\psi^n_A =\pm E^n_{\pm}\psi_B^n$, and $B\psi^n_B = \pm E^n_{\pm}\psi^n_A$. And, $(BA)\psi^n_A =(E^n_{\pm})^2\psi^n_A$, where $E^n_{\pm}=\pm \sqrt[4]{f(1-f)}e^{i\gamma_n}$. The corresponding eigenvectors are 
         \begin{eqnarray}
       \label{Eq:highdim_state}
        |E^n_{\pm}\rangle = \sum_{i=1}^{m} (\psi_A^n)_i\Big[|f\rangle_i  \pm a_i^n|1-f\rangle_i \Big],
    \end{eqnarray}
    where we define $a_i^n=(\psi_B^n)_i/(\psi_A^n)_i=|a|e^{i\phi_i^n}$. 
    
\item \textit{Point degeneracy:} Point degeneracy corresponds to $\gamma_n=\gamma$. Here, we obtain $A=|a|^2e^{2i\gamma}B^\dagger$. So, $A$ and $B$ become simultaneously diagonalizable, and we obtain $A\psi^n_A=e^{i(\phi^n +\gamma)}\sqrt{f}\psi^n_A$, and $\psi_B^n=e^{i\phi^n}|a|\psi_A^n$.  $\phi_i^n$ becomes independent of the basis $i$.  Hence $|E^n_{\pm}\rangle$ are to be treated as $m$-fold degenerate states of $H$. Now redefining $|f\rangle_n'=\sum_{i=1}^{m} (\psi_A^n)_i|f\rangle_i$, and the same for $|1-f\rangle_n'$, we obtain, from Eq.~\eqref{Eq:highdim_state}, the same equation as Eq.~\eqref{Eq:Heigenvec} up to an overall biorthogonal normalization factor of $1/\sqrt{2|a|}$ for each energy level $n$. (Refer to the Supplemental Material~\cite{Supp} for further discussion)

Interestingly, the point degeneracy situation can be achieved by the spectral flattening method from a Hamiltonian if its energies do not cross each other. In such case, we can topologically deform the Hamiltonian to a `flattened' Hamiltonian with all its eigenvalues collapsing to $\pm |E|e^{i\gamma}$, while the eigenvectors remain the same as the original ones~\cite{Kawabata_2019_PRX}. This holds as long as the energy gap is not closed in the smooth deformation process. For such a `flattened' Hamiltonian, Eq.~\eqref{Eq:h hdag one} holds. A more specialized case arises when $f=1/2$, which makes the $F$ operator $2N$-fold degenerate. This is achieved by setting the flattened Hamiltonian to have the degenerate eigenvalue with $|E|=1/\sqrt{2}$. Here, the flattened Hamiltonian becomes a normal operator. 

\end{enumerate}

For an odd-dimensional case, say $2N+1$-dimensional Hamiltonian, there are odd numbers (say $2N'+1$) of computational states at $f=\frac{1}{2}$, and the remaining $2(N-N')$ states have $f$ and $1-f$ pairs of states as in the even-dimensional case. For simplicity, we assume $N'=0$, so there is one state at the normal point $f=\frac{1}{2}$, and the remaining $2N$ states are `particle-hole' symmetric. The `particle hole' symmetric states follow the above prescription, while $|\frac{1}{2}\rangle$ state is an eigenstate $H|\frac{1}{2}\rangle=\frac{e^{i\gamma_0}}{\sqrt{2}}|\frac{1}{2}\rangle$. So, for an odd-dimensional Hamiltonian, we predict that there must be an odd number of energy states with energy amplitudes being completely \textit{`flat'} in the computational space, i.e., they only depend on the scaling parameter $d$.

\section{Discussion}\label{Sec:Discussion}

\subsection{Difficulties with NH Hamiltonian formalism}\label{Subsec:Difficulty with NH}

The necessity of the $H^{\dagger}$ operator in quantum theory is rooted in the definition of the dual space as an adjoint map. The adjoint dual space consequently dictates the self-adjoint constraint on physical operators. The associated linear algebra promotes gauge redundancy, and the charge conservation is a consequence of the global gauge symmetry. The invariants and unitarity of the theory stem from this self-adjoint condition.

In the NH counterpart, the adjoint operation does not map to the dual space. The eigenvector of $H^{\dagger}$ (equivalently, the left eigenvector of $H$) can be taken as the dual vector to the (right) eigenvector of $H$. However, given that $H\ne H^{\dagger}$, these two eigenvectors appear to belong to disjoint vector spaces without a unique map between them. Adding distinct arbitrary phases to the left and right eigenvectors, which are allowed by linear algebra, does not keep the inner product invariant. Similarly, the absence of a map between the eigenvector and dual vector hinders the unique definition of the physical operator. 

We can express an NH Hamiltonian on the basis of the Hermitian generators of some Lie Algebra (such as Pauli matrices) but with complex parameters. In other words, the parameter space of a NH Hamiltonian generally forms a complex manifold. The definitions of `length' and `angle' in a complex manifold, which are tied to the inner product of the vector space, are different from those in the real-valued manifold of Hermitian Hamiltonians. In fact, the study of quantum theory in the complex manifold is less explored in the literature, except for some efforts focused on the Hamiltonian-specific redefinitions of the Lie Algebra~\cite{SU11}, or inner products~\cite{BenderReview,MostafazadehReview}.

Moreover, the definition of the directional derivative, which gives the equation of motion, on the complex manifold does not generally coincide with the Schr\"odinger or Heisenberg pictures. On the one hand, one requires a theory with an invariant inner product on the complex manifold, while on the other hand, the theory is non-conserving by construction. This poses an apparent paradox. 

The diagonalizability of an NH Hamiltonian is, in general, violated somewhere in the parameter space of the Hamiltonian, which is called the exceptional point. The exceptional point is a physically accessible region of the parameter space where novel and functional properties can be achieved~\cite{EPBerry, EPExp, EPLee, EPLaser, EPLaser2, EPHolographic, EPMagnon, EPExp2021}, but its study is hampered due to its theoretical challenges.

\subsection{Summary of our work}\label{Subsec:Summary}
Our formalism provides precise definitions for various quantities in random NH Hamiltonians. It is based on constructing a well-defined Hilbert space of a related Hermitian operator as the computational space. In the standard second quantization formula, the Fock space is constructed from a fixed number operator or angular momentum/spin operator for any Hamiltonian. But here the computational basis is defined by the operator $H^{\dagger}H$, hence it involves a subset of the Hamiltonian's parameter space. The computational space spans between the exceptional points of the Hamiltonian, avoiding disruption by these singularities. $H^{\dagger}$ and $H$ serve as ladder operators between the states in the computational basis across the normal point, while exceptional points pose as vacua on both sides of the computational space. This makes our general formalism completely free from the singularity of exceptional points.   

 We expand the energy eigenstates on this computational basis $-$ a Bloch sphere $-$ with two expansion parameters $\theta$ and $\phi$. $0\le\phi\le 2\pi$  and $0<\theta<\pi$, with $\theta=0$ and $\pi$ are the exceptional points. The two energy eigenstates collapse into each other at exceptional points. Moreover, we discover that the dual space also lies in the same computational space, at the two antipodal points, and hence can be \textit{uniquely mapped} by discrete (space-time) transformation. In other words, we discover the existence of a computational space, which is a union of the two vector spaces of the biorthogonal eigenvectors and the exceptional points are the north and south poles (along the two bases), and the normal point lies at the equator. 

The complex expansion parameter $a\in \mathbb{S}^2$ spans between the exceptional points at the north and south poles where $a$ becomes $0$ or $\infty$. The eigenvectors and the dual vectors live on different hemispheres (separated by the equator corresponding to the normal point), and there is a gauge obstruction between them. The quantum metric involves additional (dynamical) symmetry, which we refer to as the $\mathcal{C}$ - symmetry, in analogy with the \PT-symmetry literature, to define an inner product in Sec.~\ref{Sec:Maps Dual Sym}. This suggests that the base manifold of the energy eigenstates is a non-Riemannian manifold.

We also found that the `space-time' transformations $\mathcal{S}_i$ were not necessarily the symmetry of the Hamiltonian unless the energy eigenvalues were either real or purely imaginary. In the case of real-eigenvalues, the `space-time' symmetry operators defined in the computational basis may or may not turn out to be a \PT~symmetry in the original Hamiltonian basis. Therefore, the \PT~symmetry is not a necessary condition. In \PT~symmetric Hamiltonians, generally, we have balanced gain and loss terms in the \PT~conjugate basis states. For a more general Hamiltonian with complex parameters, gain and loss terms (imaginary parameters) are not generally defined. However, our formalism always defines a `space-time' transformation in the computational basis, becoming a symmetry if the energy eigenvalues are real or purely imaginary. This does not necessarily rely on balanced gain and loss terms.  

We extended our formalism to higher dimensions but restricted ourselves to those Hamiltonians whose energy spectrum can be flattened to a circular- or point-degenerate `particle-hole' pair. This is applicable to all the Hamiltonians whose energy levels do not cross each other. For an odd-dimensional Hilbert space, the formalism predicts an odd number of the flat-energy amplitude eigenvalues, which only depend on the scaling parameter $d$, and not on any other Hamiltonian parameter. Consideration of a more general higher dimensional Hamiltonian will produce each `particle-hole' pair to correspond to a different computational basis. 
We will consider this case in a future study. 

Some similarities and differences between our formalism and the initial parts of the supersymmetry (SUSY) formalism are to be noted here. If we replace $H\rightarrow Q$ - the supersymmetric operator, then the Lie superalgebra $\{Q,Q^{\dagger}\}=H-$ the supersymmetric Hamiltonian,  coincides with Eq.~\eqref{Eq:h hdag one} where $D\rightarrow H$.
However, in the SUSY formalism, one studies the Hilbert space of $H$, which is $D$ here, and considers the Fock space of bosonic and fermionic number operators. But here we are interested in the Hilbert space of NH operator $Q$ and consider the `Fock' space of $F=Q^{\dagger}Q$. Therefore the computational space and non-Hermitian space do not possess supersymmetry.

\subsection{Applications}\label{Subsec:Applications}

In open quantum systems and quantum information, a state of interest evolves in time in an environment that causes decoherence and depolarization. There are multiple ways to study this problem, and one can deduce an effective NH Hamiltonian accordingly. For example, from the Lindblad master equation, by treating the environment as a Markovian bath of oscillators, one can define an effective NH Hamiltonian $H_{\rm eff}=H_s+i\hat{N}$, where $H_s$ is the Hermitian system Hamiltonian and $\hat{N}$ is the (Hermitian) oscillator number operator of the bath~\cite{Feshbach,Rotter,NHQM,EntanglementHam}. Alternatively, starting with the full Hamiltonian of the system + environment, one can project to an effective NH Hamiltonian $H_{\rm eff}=H_s+\Sigma$ for the system, where $\Sigma$ is the complex `self-energy' obtained by integrating out the environment's degrees of freedom~\cite{SelfenergyEmil,SelfenergyNagaosa}. The method is called the Freshbah projection method in the atomic-molecular-optical (AMO) field and is characteristically similar to Dyson's formula in the many-body quantum theory and is recently applied in many-body theory~\cite{Feshbach,Rotter,SelfenergyEmil,SelfenergyNagaosa}.
If such an effective Hamiltonian can be solved uniquely, we achieve a more transparent and solvable approach than its parent master equation for the density matrix or Green function method for the many-body counterpart.  

In photonics, the solutions of the wave equation often render NH operators in the linearized differential equations~\cite{PhotonicsRuschhaupt,PhotonicsRaghu,PhotonicsCerjan,PhotonicsCarlo}. In the cases where the solutions of the different equations belong to a vector space, tensor space, or modular space (fiber), one requires the definition of the dual space, which must be linearly transformed by the NH operator. Here, our formalism becomes convenient. 

In quantum field theory and many-body condensed matter, the phase transition is defined by an order parameter - which can be a complex operator (c.f. the superconducting order parameter). In a mean-field theory, the ground state of the whole system is a product state of eigenstates of such local order parameters~\cite{QFTFradkin,QFTWen,QFTAltland}. Our approach will be applicable to define the local Hilbert space of the NH order parameter. Similarly, in the matrix-product state (or in tensor network formalism) one builds the many-body ground state from local Hilbert space by projecting to some subspace (such as a Bell state of two nearby spin-1/2 operators)~\cite{MPSReview1,MPSReview2,QCWen}.

While the projection operator is introduced in the Hilbert space, alternatively, one can obtain an effective NH Hamiltonian for the subspace of our interest. Generally, projection operators are harder to implement in numerical simulations, and one may alternatively choose the eigenstate of an effective NH operator. Similarly, in scar states or fragmented Hilbert space, where the full Hilbert space of the many-body Hamiltonian has a subspace with some conserved charge (s) or dipole (s)~\cite{QScarBerry,QScarReview,QFragmentation}, and is nearby decoupled from the rest of the Hilbert space, any linear operator defined within that subspace is an NH operator, by construction. Here one can limit the computation within the subspace by an NH quantum theory. Recently, it has also been realized that the effective theory at the edge state of a topological insulator is an NH Hamiltonian theory~\cite{VishwanathNH,NHNoGo}. It is interesting to ask how the quantum anomaly of a topological field theory manifests in an NH quantum theory.

A class of otherwise Hermitian Hamiltonians produces complex energy and/or loss/gain systems due to outgoing or radiative boundary conditions. Related situations arise in topological edge states with open boundary conditions, semiconductor junctions with potential wells, barriers, and leads, and in photonics, where the boundary is a gapless medium for photons. If such problems have the NH Hamiltonian dynamics, they can be studied within our framework by incorporating the boundary/lead as complex terms within the
Hamiltonian or incorporated through the NH transfer matrix acting on the edge modes.

We note that the present formalism can be extended and applied to higher dimensional NH Hamiltonian made of chains of atoms, say, with periodic and open boundary conditions. This will provide an intrinsic explanation of the origin of the \textit{skin effect}~\cite{Skin1,Skin2,Skin3,Skin4,Skin5}.

\subsection{Future directions}\label{Subsec:Future dir}
There are two immediate studies to be undertaken based on the present work in the future. We have so far considered the static NH Schr\"odinger equation and built this eigenvector space and dual space. The next step is to consider the dynamics of the theory. The advantage of the present theory is that we only need to consider the time evolution of $H$, and no need to consider $H^{\dagger}$.  An immediate question is if the system is non-conserving by definition, and our formalism governs an apparently conserved inner product, how do the non-conserving physical properties get captured? Notice that the linear map is not a space-time independent operator, and hence they also evolve (explicitly or implicitly) in time. Therefore, the symmetry we have is a dynamical one (or pseudo-dynamical) in that its eigenvalues are not conserved in time. In the future study, we will focus on this aspect.  

We have focused here on the higher dimensional Hamiltonians which have no level crossing, such that each of the `particle-hole' pair eigenstates can be expressed in the same computational basis but at different points. For more general NH Hamiltonians, the $D$ operator will have different eigenvalues $d$, and each eigenvalue corresponds to a different computational basis. Then, the discussion in Sec. \ref{Sec:Hig dim} can be generalized to a larger Hilbert space. Consideration of more general higher-dimensional Hilbert spaces will elucidate the emergence of higher-order exceptional points. It will also bypass the numerical search~\cite{Mailybaev2006} of exceptional points with precise analytical predictions through the identification of the computational basis states.
The applications of the theory in various settings as discussed in Sec. \ref{Subsec:Applications} will subsequently be carried out.

\section*{Acknowledgments}
We thank Sachindeo Vaidya, Baladitya Suri, Girish Agarwal, Ronny Thomale, and Sourin Das for useful discussions. PB acknowledges the Prime Minister Research Fellowship (PMRF) from the Government of India for the fellowship award. TD's research is supported by the Science and Engineering Research Board (SERB) under the Department of Science and Technology (DST) of the Government of India for the CRG Research Grant CRG/2022/003412.

\appendix
\section{}

\subsection{Exceptional points of $H$ as the vacuum states of $F$}\label{Appendix:EP as vacuum states}

The exceptional points correspond to the parameter value where the Hamiltonian becomes non-diagonalizable, in the sense that the two or more eigenstates become linearly dependent on each other. In our formalism of expressing energy eigenstates in the computational basis, we uncover that at the exceptional point, which is the zero-energy state, the energy eigenstates collapse to one of the computational basis states. The computational space is parameterized by $f$, which is bounded by the exceptional points at $f=0$ and $f=1$. At the $f=0$ boundary, Eqs.~\eqref{Eq:ladder action a}, \eqref{Eq:ladder action d} give $H|0\rangle=0$ and $H^\dagger|1\rangle=0$, while Eqs.~\eqref{Eq:ladder action b}, \eqref{Eq:ladder action b} produce $H|1\rangle=e^{i(\gamma-\phi)}|0\rangle$ and $H^\dagger|0\rangle=e^{-i(\gamma-\phi)}|1\rangle$. Clearly, here $H^2=(H^{\dagger})^2=0$, and the formalism coincides with the fermionic algebra. At the other boundary $f=1$, the situation is reversed. Therefore, the two exceptional points are the vacuum states in the usual Fock space language.

\subsection{Degenerate $F$ at the normal point}\label{Appendix:NP in 2D}
The normal point occurs at $f=1/2$, where $|f\rangle$ and $|1-f\rangle$ are degenerate. Here the ladder action specified in Eq. \eqref{Eq:ladder action} no longer holds in general. Instead, the ladder operators ($H$, $H^{\dagger}$) can take anywhere in the degenerate manifold of the computational space. Here we have $F=\frac{\mathbb{I}}{2}$ (this follows from Eq. \eqref{Eq:h hdag one} which can be rewritten as $\left[H,H^\dagger\right]=\mathbb{I}-2F$).
Hence, any pair of orthogonal states is an eigenbasis of $F$. In order to make our formalism work in a similar way, it is sufficient to find a pair of orthogonal states such that the ladder action in Eq. \eqref{Eq:ladder action} continues to hold. We denote this orthogonal pair of states as $|f\rangle \equiv \left|\frac{1}{2},m\right\rangle$ and $|1-f\rangle \equiv \left|\frac{1}{2},n\right\rangle$, where $m$ and $n$ are the degeneracy labels. We will use these as our computational basis. The ladder action on this basis can be specified as,
\begin{equation}
H\left| \frac{1}{2},m\right\rangle = \frac{e^{i(\gamma+\phi)}}{\sqrt{2}} \left| \frac{1}{2},n\right\rangle, ~
H\left| \frac{1}{2},n\right\rangle = \frac{e^{i(\gamma-\phi)}} {\sqrt{2}}\left| \frac{1}{2},m\right\rangle  \label{14},
\end{equation}
and similarly for $H^\dagger$. It is now obvious that our further analysis of the non-degenerate computational basis states remains valid here. For example, for the energy eigenspectrum, we obtain $E_\pm =\pm \frac{e^{i \gamma}}{\sqrt{2}}$ and $|E_\pm \rangle =\frac{1}{\sqrt{2}}\left(\left |\frac{1}{2},m\right\rangle \pm e^{i \phi} \left |\frac{1}{2},n\right\rangle\right)$.

We emphasize here that at the normal point, where $H$ commutes with $H^\dagger$, the right and left eigenvectors become identical, i.e. $|E_n\rangle=|\tilde{E}_n\rangle$. Consequently, the dual space is defined by the usual Hermitian conjugation map. This has been confirmed earlier in Sec.~\ref{Subsec:Dual} through our metric $\mathcal{C}_1^\dagger \mathcal{U}_1$ reducing to identity at the normal point $|a|=1$. An explicit evaluation of $\langle E_n|E_m\rangle$, which equals to $\delta_{nm}$ only at $|a|=1$, also validates the same.

\subsection{Generalized Bernard LeClair classification}\label{Appendix:GBL}

In this section, we extend our discussion of the generalized Bernard LeClair (gBLC) classification of the Hamiltonian in Sec.~\ref{Subsec:Sym}\cite{BLC,BdG2NH,NHClassification1,NHClassification2,NHClassification3,VishwanathNH}. In gBLC scheme, a general NH Hamiltonian is classified based on the symmetry classes defined by Eq.~\ref{Eq:GBL}. We now discuss these four symmetry classes at length, 

\begin{enumerate}
    \item \underline{K-symmetry}: $\Xi=K$, $f(H)=H^*$, $\epsilon_K=\pm 1$ and $KK^*=\eta_K \mathbb{I}$ with $\eta_K=\pm 1$

    \item \underline{Q-symmetry}: $\Xi=Q$, $f(H)=H^\dagger$, $\epsilon_Q=\pm 1$ and $Q^2=\eta_Q\mathbb{I}$ with $\eta_Q=1$

    \item \underline{C-symmetry}: $\Xi=C$, $f(H)=H^T$, $\epsilon_C=\pm 1$ and $CC^*=\eta_C \mathbb{I}$ with $\eta_C=\pm 1$

    \item \underline{P-symmetry}: $\Xi=P$, $f(H)=H$, $\epsilon_P= -1$ and $P^2= \eta_P\mathbb{I}$ with $\eta_P=1$
\end{enumerate}

In this classification scheme, $H \longrightarrow iH$ is considered an inequivalent transformation, which is the case where a line gap exists in the energy spectrum~\cite{GBL_Liu_Chen_2019_PRB}. Further details of the equivalence relations for these symmetry classes can be found~\cite{GBL_Liu_Chen_2019_PRB}.

With these, it is interesting to note that our formalism reveals important physical properties of general NH systems (for example, topological defect classification~\cite{GBL_Liu_Chen_2019_PRB} etc.) by unveiling the connection of $H$ with different gBLC symmetry classes. We start with the chiral symmetry of $H$ with $\mathcal{Q}=\sigma_z$ as $H=-\mathcal{Q} H \mathcal{Q}^{-1}$. This immediately reveals that $H$ belongs to the symmetry P-class (with $\epsilon_{P}=\epsilon_{\mathcal{Q}} =1$) of the gBLC classification.

We now turn to our unitary and anti-unitary maps, $\mathcal{U}_i$ and $\mathcal{A}_i$, and observe the following. For the antiunitary, $\mathcal{A}_i=\mathcal{V}_i\mathcal{K}$, we find $H=\epsilon_{\mathcal{V}_i} \mathcal{V}_{i} H^T \mathcal{V}_{i}^{-1}$ with $\epsilon_{\mathcal{V}_i}=(-1)^{i+1}$  where $i=1,2$ (refer Eq. \eqref{Eq:AHHdag}). This corresponds to the C-type symmetry class with $\epsilon_C=\epsilon_{\mathcal{V}_i}$ and $\eta_C=\eta_{\mathcal{V}_i}=(-1)^{i+1} \mathbb{I}$ of the gBLC classification. Contrary to the anti-unitary maps, we observe that for complex energies, the unitary maps do not correspond to any of the gBLC classes. Instead, they correspond to a 'dynamic' version of the gBLC symmetry class-Q due to the presence of the $e^{2i\gamma}$ factor in Eq. \eqref{Eq:UHHdag}. These coincide with the static definition of the gBLC symmetry class when the energies are either real or pure imaginary as follows: (a) \textit{for real energy eigenvalues}, $e^{i\gamma}=\pm 1$, we find $H=\epsilon_{\mathcal{U}_i} \mathcal{U}_{i} H^\dag \mathcal{U}_{i}^{-1}$ with $\epsilon_{\mathcal{U}_i}=(-1)^{i+1}$, which corresponds to the symmetry class-Q with $\epsilon_Q=\epsilon_{\mathcal{U}_i}$ and $\eta_Q=\eta_{\mathcal{U}_i} =(-1)^{i+1} \mathbb{I}$ (see footnote~\footnote{In the gBLC classification, $\eta_Q=-1$ is not included because the minus sign can be taken care of by a global gauge fixing of the unitary $Q$. We have not done that for $\eta_{\mathcal{U}_2}$ because the global gauge of $\mathcal{U}_2$ in our analysis is fixed to facilitate the ease of notation for our results in Sec. \ref{Sec:Maps Dual Sym}.}) of the gBLC classification of random non-Hermitian matrices; (b) \textit{for imaginary energy}, $e^{i\gamma}=\pm i$, we find $H=-\epsilon_{\mathcal{U}_i} \mathcal{U}_{i} H^\dag \mathcal{U}_{i}^\dag$ which corresponds to the symmetry class-Q with $\epsilon_Q=-\epsilon_{\mathcal{U}_i}$.

Next, we look at the symmetry operations, $\mathcal{S}_i=\mathcal{W}_i\mathcal{K}$. For similar reasons as for the unitary maps, $\mathcal{S}_i$ do not correspond to any of gBLC classes for complex energies. For the special case of real and imaginary energy, there is an identification with the gBLC classes as follows: (i) \textit{for real energies}, we find $H=\epsilon_{\mathcal{W}_i} \mathcal{W}_{i} H^* \mathcal{W}_{i}^{-1}$ with $\epsilon_{\mathcal{W}_i}=(-1)^{i+1}$, which corresponds to the symmetry class-K with $\epsilon_K=\epsilon_{\mathcal{W}_i}$ and $\eta_K=\eta_{\mathcal{S}_i}=1$; (b) \textit{for imaginary energies}, we find $H=-\epsilon_{\mathcal{W}_i} \mathcal{W}_{i} H^* \mathcal{W}_{i}^\dag$ which corresponds to the symmetry class-K with $\epsilon_K=-\epsilon_{\mathcal{W}_i}$.

From the above discussion, we observe that $\epsilon_{\mathcal{V}_1}$ and $\epsilon_{\mathcal{V}_2}$ are not independent but are related by a minus sign. This is because the corresponding unitary operators are related as $\mathcal{V}_2=\mathcal{Q}\mathcal{V}_1$ (follows from $\mathcal{A}_2=\mathcal{Q}\mathcal{A}_1$). In the language of gBLC classes, this implies that $\mathcal{V}_2$ is a combination of a P-type ($\mathcal{Q}$) and C-type ($\mathcal{V}_1$) symmetry. The result of this combination is another $C$-symmetry with the corresponding $\epsilon_\Xi$ related as  $\epsilon_{\mathcal{V}_2}=\epsilon_{\mathcal{Q}}\epsilon_{\mathcal{V}_1}$. Similar is the case for $\epsilon_{\mathcal{U}_i}$ and $\epsilon_{\mathcal{W}_i}$ as $\mathcal{U}_2=\mathcal{Q}\mathcal{U}_1$ and $\mathcal{S}_2=-\mathcal{Q}\mathcal{S}_1$, respectively.

It can be checked that a combination of C-type and Q-type symmetry classes as $\left(C^{-1}Q\right)^*$, results in a K-type symmetry with $\epsilon_{K}=\epsilon_{Q}\epsilon_{C}$. This holds true for our symmetry operations $\mathcal{S}_i$ as well since $\mathcal{W}_1=\left(\mathcal{V}_1^{-1}\mathcal{U}_1\right)^* =\left(\mathcal{V}_2^{-1}\mathcal{U}_2\right)^*$ (follows from $\mathcal{S}_1=\mathcal{A}_1^{-1}\mathcal{U}_1 = \mathcal{A}_2^{-1}\mathcal{U}_2$) and $\mathcal{W}_2=\left(\mathcal{V}_1^{-1}\mathcal{U}_2\right)^* =\left(\mathcal{V}_2^{-1}\mathcal{U}_1\right)^*$.

\section{More examples}\label{Appendix:MoreExamples}

\subsection{A generic two-parameter Hamiltonian}\label{Appendix:TwoparameterH}
\begin{figure}[ht]
  \centering
  \includegraphics[width=7cm]{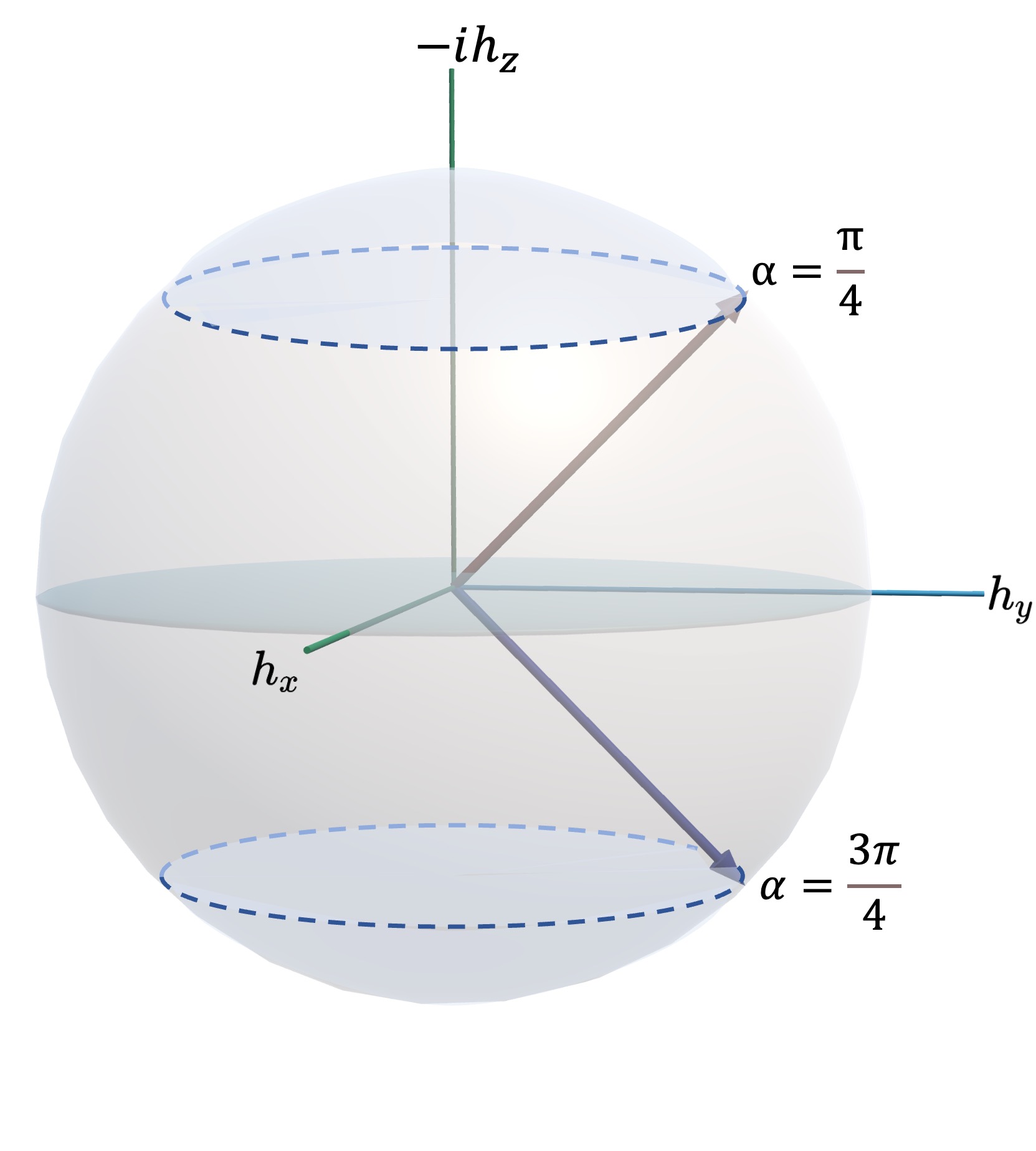}
  \caption{The parameter space of the Hamiltonian defined in Sec.~\ref{Subsec:Example alpha beta} gives a sphere of radius $d=2|h|^2=1$.  The $\alpha=\pi/4$ and $3\pi/4$ circles give exceptional contours while $\alpha=\pi/2$ is the normal contour.  The energy eigenvalues are real (imaginary) within (outside) the regions bounded by the exceptional points/contours. In the region with real eigenvalues, $\mathcal{S}_1$ becomes a static symmetry and coincides with the \PT-symmetry. }
  \label{Fig:Example alpha beta}
\end{figure}

It is instructive to demonstrate the above formalism and the symmetry-enforced real eigenvalues and symmetry-breaking phase transition with a few Hamiltonian parameters. Here we consider a NH Hamiltonian with $h_x=\frac{1}{\sqrt{2}}\sin\alpha\cos\beta$, $h_y=\frac{1}{\sqrt{2}}\sin\alpha\sin\beta$, and $h_z=i\frac{1}{\sqrt{2}}\cos\alpha$, where $0\le \alpha\le\pi$, $0\le \beta\le 2\pi$. This gives $d=2|h|^2=1$. The Hamiltonian parameter space is shown in Fig. \ref{Fig:Example alpha beta}.


For the computational space we have $\mathfrak{f}_x=-\frac{1}{2}\sin2\alpha\sin\beta$, and $\mathfrak{f}_y=\frac{1}{2}\sin2\alpha\cos\beta$, $\mathfrak{f}_z=0$. So the computational basis lives on a circle $\mathbb{S}_{\mathfrak{f}}^1$ defined by $|\mathfrak{f}|=\frac{1}{2}|\sin 2\alpha|$, as shown in Fig. \ref{Fig:H param space}. The eigenvectors are 

\begin{align}\label{Eq:f 1-f example vec}
\left|f\right\rangle, \left|1-f\right\rangle =\frac{1}{\sqrt{2}}\left(\begin{array}{c}
1 \\
\pm i\nu_{\alpha} e^{i\beta}
\end{array}\right)
\end{align}
where $\nu_{\alpha}=\text{sgn}\left[\sin 2\alpha\right]$. The corresponding eigenvalues are $f=\frac{1}{2}+|\mathfrak{f}|$, and $1-f=\frac{1}{2}-|\mathfrak{f}|$. The exceptional contours are defined at the roots of $f$ and $1-f$, which lie at 
$\alpha_n=\frac{\pi}{4}(2n\pm 1)$ for $n\in \mathbb{Z}$, and the normal contour lies at $\alpha=\frac{n\pi}{2}$ for $n \in \mathbb{Z}$, as shown in Fig. \ref{Fig:Example alpha beta}.

We go from the Hamiltonian to the computation basis by the unitary transformation $U=(|f\rangle ~ |1-f\rangle)$. In the computational basis, the Hamiltonian is off-diagonal as in Eq. \eqref{Eq:HinF}, with $\sqrt{f}e^{i\left(\gamma+\phi\right)}=i\sin\left(\frac{\pi}{4}+\nu_{\alpha}\alpha\right)$ and $\sqrt{1-f}e^{i\left(\gamma-\phi\right)}=i\sin\left(\frac{\pi}{4}-\nu_{\alpha}\alpha\right)$. Interestingly, the Hamiltonian in the computational basis has one less parameter due to the constraint $\{H,H^{\dagger}\}=\mathbb{I}$. We note that the two phases $\gamma$ and $\phi$ are only defined to be $\frac{m\pi}{2}$, $m \in \mathbb{Z}$ (i.e. 
$E_{\pm}$ and $a$ are either real or imaginary here). The Hamiltonian parameter space $\mathbb{S}^2_h$ splits into three regions by the exceptional points ($\alpha_n$): 

\begin{enumerate}
    \item\textit{Imaginary $E_{\pm}$}: $\alpha \in \left(0,\frac{\pi}{4}\right)$,  $\phi=0$, and $\gamma=\frac{\pi}{2}$.
    \item\textit{Real $E_{\pm}$}: 
    \begin{enumerate}
        \item $\alpha \in \left(\frac{\pi}{4},\frac{\pi}{2}\right)$,  $\phi=\frac{3\pi}{2}$, and $\gamma=\pi$.
        \item $\alpha \in \left(\frac{\pi}{2},\frac{3\pi}{4}\right)$,  $\phi=\frac{\pi}{2}$, and $\gamma=\pi$.
    \end{enumerate}
    \item\textit{Imaginary $E_{\pm}$}: $\alpha \in \left(\frac{3\pi}{4},\pi\right)$, $\phi=\pi$, and $\gamma=\frac{\pi}{2}$. 
\end{enumerate}

For fixed values of $\phi=\frac{m\pi}{2}$, the dual space maps $\mathcal{U}_{1,2}$ become static. $\mathcal{S}_{1,2}$ become the symmetry/anti-symmetry of the Hamiltonian since $\gamma=\frac{n\pi}{2}$, $n \in \mathbb{Z}$, i.e., energies are either real or imaginary. We discuss their representations in the three cases. 

In region (2a), $\alpha\in\left(\frac{\pi}{4},\frac{\pi}{2}\right)$, where energies are \textit{real}: $\mathcal{U}_1=-\sigma_y$, $\mathcal{U}_2=i\sigma_x$ and $\mathcal{S}_1=i\sigma_z\mathcal{K}$, and $\mathcal{S}_2=-i\mathcal{K}$, with $\mathcal{S}_1$ is the symmetry of the Hamiltonian. If we revert to the Hamiltonian basis, we obtain $\mathcal{S}_1=e^{i\beta}\tau_x\mathcal{K}$, where $\tau_i$ are the $2\times 2$ Pauli matrices in the Hamiltonian basis, and $\mathcal{S}_2=-\frac{i}{2}\left[(1-e^{2i\beta})\mathbb{I}+(1+e^{2i\beta})\tau_z\right])\mathcal{K}$. In this basis $\mathcal{S}_2$ becomes a dynamical symmetry operation, while $\mathcal{S}_1$ becomes precisely the static \PT~symmetry operator (upto a global gauge of $e^{i\beta}$) introduced by Bender et al~\cite{BenderReview}. The $\mathcal{C}_1$ operator defined in Eq.\eqref{Eq:C1 in F} remains dynamical due to the presence of $|a|$ term. In region 2(b), on the other hand, the symmetry operators $\mathcal{S}_i$ are found to differ by a phase of $\pi$ compared to those in region 2(a).

In region (1), where the energies are purely imaginary, $\mathcal{S}_i$ changes to a different form as $\mathcal{S}_1=\mathcal{K}$ and $\mathcal{S}_2=-\sigma_z\mathcal{K}$. In the Hamiltonian basis $\mathcal{S}_{1}=\frac{1}{2}\left[(1-e^{2i\beta})\mathbb{I}+(1+e^{2i\beta})\tau_z\right]\mathcal{K}$ and $\mathcal{S}_2=ie^{i\beta}\tau_x\mathcal{K}$, which become dynamical and static, respectively. The operators $\mathcal{S}_i$ in the region (3) acquire an additional phase of $\pi$ relative to the corresponding operators in region (1).

A Hermitian condition is achieved at $\alpha=\pi/2$, which is one of the normal surfaces lying inside the real energy region. Here we get $E_\pm=\mp\frac{1}{\sqrt{2}}$ with $\gamma=\pi$. $|a|=1$ at the normal point and we have the freedom to choose $a=-i$. This gives $\mathcal{U}_1=-\sigma_y$, $\mathcal{U}_2=i\sigma_x$ and $\mathcal{C}_1=-\sigma_y$, with $\mathcal{C}_1^{\dagger}\mathcal{U}_1=\mathbb{I}$ and $\mathcal{C}_2^{\dagger}\mathcal{U}_2=\mathcal{Q}$. Then the inner product definition in Eq. \eqref{Eq:Pos norm U1C1} coincides with the usual Hermitian conjugation dual space. It  is not surprising that the Hermitian condition is achieved in the region (2) with real eigenvalues. Because, for any NH Hamiltonian with real eigenvalues, one can find a similarity transformation (smooth deformation) to obtain a Hermitian Hamiltonian, which lies at the normal point. Clearly, the smooth deformation is defined by the $\alpha$ parameter (see Supplemental Material~\cite{Supp} for more examples).

\subsection{A chiral (Hermition or NH) Hamiltonian}\label{Appendix:Chiral}

We consider a class of traceless Hamiltonians $h$ defined as 
\begin{align}
h = \left(\begin{array}{cc}
d_3 & d_1 H \\
d_2 H^{\dagger} & -d_3
\end{array}\right),
\label{chiralHam}
\end{align}
in which $\{H,H^{\dag}\}=\mathbb{I}$, and $d_i\in \mathbb{C}$. As a special case, we consider a chiral Hermitian Hamiltonian when $d_3=0$ and $d_1=d_2^{\dagger}=1$, while the formalism can be easily generalized to the NH case. (This is precisely the chiral operator considered in the work of Feinberg and Zee~\cite{JoshuaZee} discussed in the Introduction (Sec.~\ref{Sec:Introduction})). $h$ anti-commutes with the unitary chiral operator $\mathcal{Q}=\sigma_z\otimes I$, where $I$ is a unit matrix of the same dimension as $H$. We encounter such Hamiltonians in quantum condensed matter physics, which has low-energy Dirac dispersion. Such Hermitian chiral Hamiltonians also find importance in the context of non-Hermitian systems as any NH Hamiltonian with a point gap can be mapped to such a Hamiltonian with similarity transformation.  The chiral symmetry ensures that all energy eigenvalues $e$ come in particle-hole pair $\pm e$. Now solving for the eigenvalue of $h$: $h|e\rangle=e|e\rangle$, we get $e_1^{\pm}=\pm \sqrt{f}$ and $e_2^{\pm}=\pm \sqrt{1-f}$ as the four eigenvalues. For the eigenvectors, we get spinors of the form, 
\begin{align}
\centering
\left|e_1^{\pm}\right\rangle &=\frac{1}{\sqrt{2}}\left(\begin{array}{c}
|1-f\rangle \\
\pm e^{-i(\phi+\gamma)}|f\rangle
\end{array}\right),\\ 
\left|e_2^{\pm}\right\rangle &=\frac{1}{\sqrt{2}}\left(\begin{array}{c}
|f\rangle \\
\pm e^{i(\phi-\gamma)}|1-f\rangle
\end{array}\right).
\end{align}

\begin{figure}[t]
    \includegraphics[width=8cm]{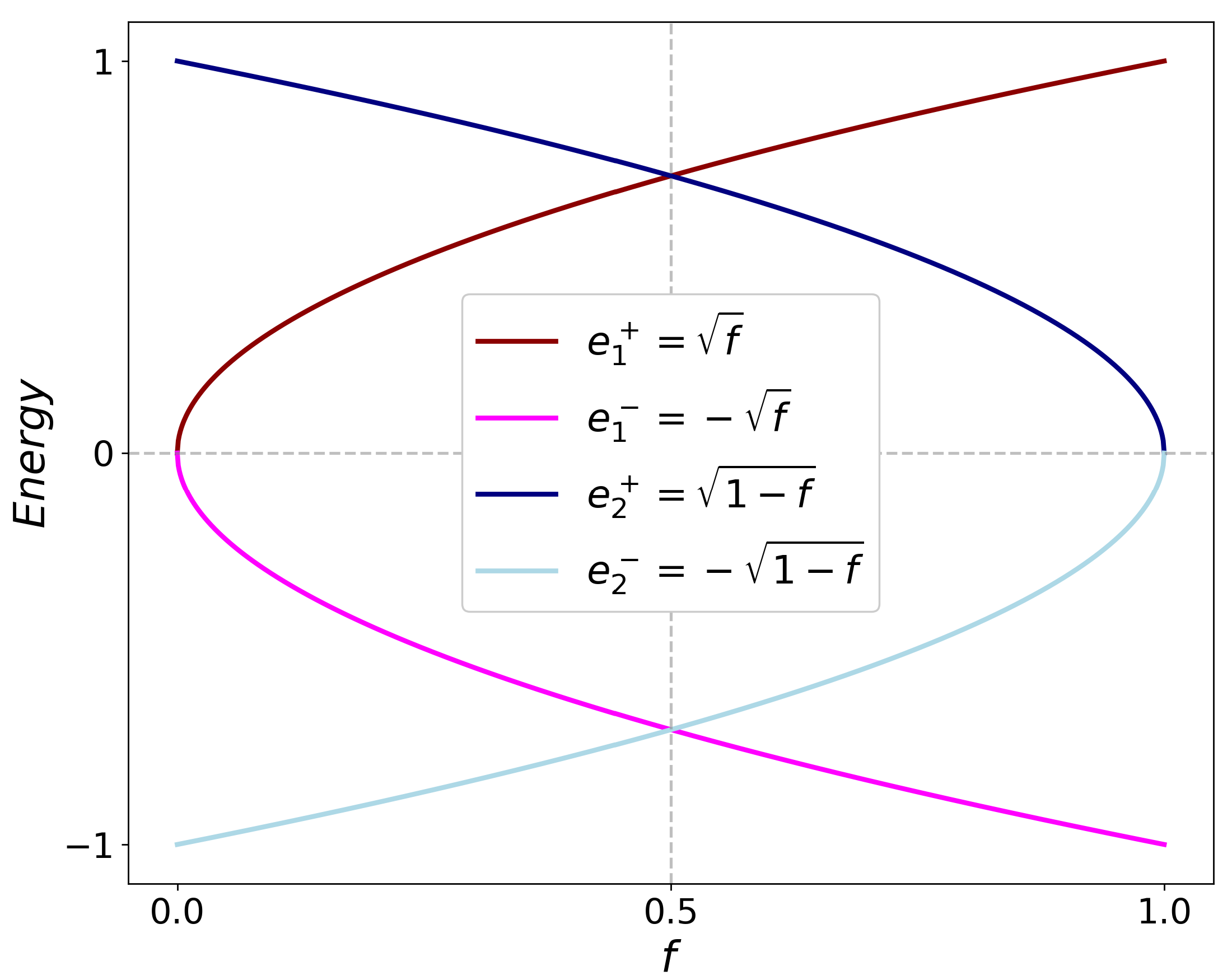}
    \caption{Four energy levels of the chiral Hermitian Hamiltonian in Eq.~\eqref{chiralHam} are plotted as a function of $f$. We find that at the exceptional points for $H$, i.e., at $f=0,1$, we have a  degeneracy in $h$ at $e_{1,2}=0$ between the eigenstates $|e_1^{\pm}\rangle$ and $|e_2^{\pm}\rangle$, respectively. On the contrary, at the normal point of $H$, $h$ also has a degeneracy but at finite energies where level crossings occur between the eigenstates $e_{1,2}^{+}$ and $e_{1,2}^{-}$.}
    \label{Fig:Chiral Hamiltonian}
\end{figure}

We observe that the exceptional and normal points in $H$ translate into interesting level crossings, and degeneracies in the Hilbert space of the chiral Hermitian Hamiltonian, $h$ (see Fig. \ref{Fig:Chiral Hamiltonian}). In particular, we observe two two-fold degeneracies at the exceptional points: (i) at $f=0$ the states $|e_1^{\pm}\rangle$ become degenerate at $e_1=0$, while (ii) at $f=1$ the states $|e_2^{\pm}\rangle$ become degenerate at $e_2=0$. At the normal point ($f=\frac{1}{2}$), on the other hand, we observe two level crossings as follows: (i) $|e_{1,2}^{+}\rangle$ become degenerate pair with energy $e_{1,2}^+=\frac{1}{\sqrt{2}}$, and (ii) $|e_{1,2}^{-}\rangle$ with energy $e_{1,2}^-=-\frac{1}{\sqrt{2}}$.

\subsection{$4D$ - Hilbert space}\label{Appendix:4D}
Here we consider an example of a 4-dimensional NH Hamiltonian. To make the discussion easy, we consider the cases where the $D=\{H,H^{\dagger}\}$ operator is four-fold degenerate. This is obtained when $H$ is expandable in terms of five $\Gamma_a$-matrices which follow $\{\Gamma_{\mu},\Gamma_{\nu}\}=2\delta_{\mu\nu}$, $\mu,\nu=1-5$. A set of such $\Gamma_{\mu}$ matrices is $\Gamma_{1,2}=\tau_3\otimes \sigma_{1,2}$,  $\Gamma_{3,4}=\tau_{1,2}\otimes \sigma_0$, and $\Gamma_5=\tau_3\otimes\sigma_3$, where $\tau_{\mu}$ and $\sigma_{\mu}$ are $2\times 2$ Pauli matrices and $\sigma_0$ is $2\times 2$ identity matrix. Expressing a traceless Hamiltonian as $H=\sum_{\mu=1}^5 h_{\mu}\Gamma_{\mu}$, where $h_{\mu}\in \mathbb{C}$, we find $D=d\mathbb{I}_{4\times 4}$, where $d=2|h|^2=1$ (set), and $F=\left[\frac{\mathbb{I}}{2}+i\sum_{\mu,\nu>\mu}\mathfrak{f}_{\mu\nu}\Gamma_{\mu}\Gamma_{\nu}\right]$, where $\mathfrak{f}_{\mu\nu}=2{\rm Im}(h_{\mu}^*h_{\nu})\in\mathbb{R}$ as in Sec.~\ref{Subsec:General Example}. 

We take the same example as in Sec.~\ref{Subsec:Example alpha beta} that $h_1=\frac{1}{\sqrt{2}}\sin\alpha\cos\beta$, $h_2=\frac{1}{\sqrt{2}}\sin\alpha\sin\beta$, and $h_3=i\frac{1}{\sqrt{2}}\cos\alpha$, where $0\le \alpha\le\pi$, $0\le \beta\le 2\pi$. Then the only surviving components of $\mathfrak{f}_{\mu\nu}$ are $\mathfrak{f}_{13}=\frac{1}{2}\sin 2\alpha \cos\beta$ and $\mathfrak{f}_{23}=\frac{1}{2}\sin 2\alpha \sin\beta$ and we define $|\mathfrak{f}|=\sqrt{\mathfrak{f}_{13}^2+\mathfrak{f}_{23}^2}=\frac{1}{2}|\sin2\alpha|$. This gives $f=\frac{1}{2}+|\mathfrak{f}|$ and $1-f=\frac{1}{2}-|\mathfrak{f}|$ and the corresponding degenerate eigenvectors are   
\begin{equation}
\left|f\right\rangle_1,\, \left|1-f\right\rangle_1 =\frac{1}{\sqrt{2}}
\begin{pmatrix}
\pm i\nu_{\alpha} e^{-i\beta} \\
0\\
0\\
1
\end{pmatrix},
\end{equation}

\begin{equation}
\left|f\right\rangle_2,\, \left|1-f\right\rangle_2 =\frac{1}{\sqrt{2}}
\begin{pmatrix}
0 \\
\pm i\nu_{\alpha} e^{i\beta}\\
1\\
0
\end{pmatrix}, 
\end{equation}
where $\nu_{\alpha}=\text{sgn}\left[\sin 2\alpha\right]$. In this computational basis, the Hamiltonian becomes off-block diagonal $H=\begin{pmatrix}
0 & B\\
A & 0
\end{pmatrix}$, where $A,B=-\sin\left[\alpha\pm\nu_{\alpha}\pi/4\right]\left[\cos\beta\sigma_x -\sin\beta\sigma_y \right]$, both being Hermitian. Equating 
$A/B=|a|^2e^{2i\gamma}$, we get $e^{i\gamma}=i\sqrt{{\rm sgn}[\cos 2\alpha]}$. Therefore $\gamma$ is independent of the energy level index and hence we have a point degeneracy. Here  $\psi_A^n$ are eigenstates of $A$ with eigenvalues $\sqrt{f}e^{i(\gamma+\phi^n)}$ given by  
\begin{equation}
\psi_A^{1,2}
=\frac{1}{\sqrt{2}}
\begin{pmatrix}
1\\
\pm e^{-i\beta}
\end{pmatrix}.
\end{equation}
This gives $\phi^n=-\gamma+n\pi$,  $\phi^n_i=\phi^n$, and $\psi_B^n=|a|e^{i\phi^n}\psi_A^n$. Substituting them in Eq.~\eqref{Eq:highdim_state}, we obtain the degenerate energy eigenstates $|E_{\pm}^{1,2}\rangle$ with the eigenvalues $E^1_{\pm}=E^2_{\pm}=\pm i\sqrt{\cos{2\alpha}/2}$.

\subsection{$3D$-Hilbert space}\label{Appendix:3D}
To demonstrate the existence of flat-amplitude energy level, we consider an example of a 3-dimensional NH Hamiltonian. Such a Hamiltonian may be suitably expressed in the three-dimensional representation of Pauli matrices, or in a subset of Gell-Mann Matrices which follow  $\{\gamma_{\mu},\gamma_{\nu}\}=2\delta_{\mu,\nu}$ such as $D$ operator is three-fold degenerate. We consider an example 
\begin{equation}
    H=\frac{1}{2\sqrt{2}}
    \begin{pmatrix}
        0 & 2\cos \kappa & 2\cos \kappa\\
        2\sin \kappa & 1 & - 1\\
        2\sin \kappa & -1 & 1
    \end{pmatrix},
\end{equation}
where $0\le\kappa\le2\pi$. The Hamiltonian is suitably normalized such that  $\left\{H,H^\dag\right\}=\mathbb{I}$. Expectedly, one of the eigenvalues of $F$ lies at the normal point  i.e., $f_0=1/2$, while the other two are at $f=\sin^2\kappa$ and $1-f=\cos^2\kappa$. The corresponding eigenvectors of $F$ are,
\begin{equation}
    \left|\frac{1}{2}\right\rangle =\frac{1}{\sqrt{2}}
    \begin{pmatrix}
        0\\
        1\\
        -1
    \end{pmatrix}, 
    |f\rangle =
    \begin{pmatrix}
        1\\
        0\\
        0
    \end{pmatrix}, 
    |1-f\rangle =\frac{1}{\sqrt{2}}
    \begin{pmatrix}
        0\\
        1\\
        1
    \end{pmatrix}.
\end{equation}
As discussed in Sec. \ref{Sec:Hig dim}, $\left|\frac{1}{2}\right\rangle$ is  the simultaneous eigenstate of $F$  and $H$ with $e^{i\gamma_0}=1$ for this Hamiltonian. For $|f\rangle$ and $|1-f\rangle$ states, we use Eq. \ref{Eq:h hdag one} to find $\sqrt{f}e^{i(\gamma+\phi)}=\sin\kappa$ and $\sqrt{1-f}e^{i(\gamma-\phi)}=\cos\kappa$.  Substituting them we get the particle-hole eigenvalues to be $E_{\pm}=\sqrt{\sin2\kappa/2}$, with  $e^{i\gamma}=\sqrt{\text{sgn}(\sin2\kappa)}$. The expansion coefficient is $a=\sqrt{\tan\kappa}$ with $e^{i\phi}=\sqrt{\text{sgn}(\tan\kappa)}$.




 \bibliography{tempbib}

\pagebreak
\begin{center}
\textbf{\large Supplemental Material: A Hermitian bypass to the non-Hermitian quantum theory}
\end{center}
\setcounter{equation}{0}
\setcounter{figure}{0}
\setcounter{table}{0}
\setcounter{page}{1}
\setcounter{section}{0}
\makeatletter
\renewcommand{\theequation}{S\arabic{equation}}
\renewcommand{\thesection}{S \Roman{section}}
\renewcommand{\thefigure}{S\arabic{figure}}
\renewcommand{\bibnumfmt}[1]{[S#1]}
\renewcommand{\citenumfont}[1]{S#1}

\section{Generalization to $d \neq 1$}
We now generalize to $ \left\{H,H^\dagger\right\}=d\mathbb{I} $ with $ d \geq 0$ (and $d$ is not set equal to $1$). It is obvious that compared to $ \left\{H',H'^\dagger\right\}=\mathbb{I} $ where $H'=H/\sqrt{d}$, this will only scale the eigenvalues (and the magnitudes of the prefactors in the ladder action) and will leave the computational and the biorthogonal basis unchanged. For instance, for the computational space, we find,
\begin{align}
    F|f\rangle &= f |f\rangle , \nonumber\\
    F|d-f\rangle &= (d-f) |1-f\rangle ,
\end{align}
where $0 \leq f \leq d$, and $|f\rangle$ and $|d-f\rangle$ are also the eigenstates of operator $F'=H'^\dagger H'$ with eigenvalues $f'=f/d$ and $(1-f')=(d-f)/d$, respectively. We note here that the computational spectrum displays the same features as before. For example, the boundary of the computational spectrum, $f=0,d\,$, corresponds to the exceptional surfaces whereas the center, $f=d/2$, corresponds to the NP. The amplitudes of the prefactors in the cyclic ladder action are modified as below, 
\begin{subequations}\label{Supp_Eq:ladder action}
\begin{align}
H|f\rangle &= e^{i(\gamma+\phi)} \sqrt{f}|d-f\rangle; \label{Supp_Eq:ladder action a}\\
H|d-f\rangle &= e^{i(\gamma-\phi)} \sqrt{d-f}|f\rangle  \label{Supp_Eq:ladder action b}\\
H^{\dagger}|f\rangle &= e^{-i(\gamma-\phi)} \sqrt{d-f}|d-f\rangle, \label{Supp_Eq:ladder actionc}\\
H^{\dagger}|d-f\rangle &= e^{-i(\gamma+\phi)} \sqrt{f}|f\rangle, \label{Supp_Eq:ladder action d}
\end{align}
\end{subequations}
which specifies the energy space with $E_{\pm}=\pm e^{i\gamma} |E|$ where $|E|=\sqrt[4]{f(d-f)}$ and the corresponding eigenvectors,
\begin{equation}
    |E\rangle_{\pm}=\frac{1}{\sqrt{2|a|}}\left(|f\rangle \pm e^{i\phi}|a||d-f\rangle\right),\label{Supp_h1 spectrum}
\end{equation}
where $a=|a|e^{i\phi}$ and $|a|=\sqrt[4]{f/(d-f)}$.



\section{Equivalence of $\left\{H,H^\dagger\right\}=d\mathbb{I}$ and $\left[H,D\right]=0$}
In this section, we note that the restriction to a single eigenvalue of $D$ is equivalent to constraing to a set of operators $D$ that commute with the Hamiltonian. We first observe that $\left[H,D\right]=0$ implies $\left[H^\dagger,D\right]=0$ and $\left[F,D\right]=0$. This gives,
\begin{align}
    F\left(H|f\rangle\right) &= DH|f\rangle -HF|f\rangle, \nonumber\\
    &=HD|f\rangle -HF|f\rangle,
\end{align}
Since $\left[F,D\right]=0$, and $F$ and $D$ are both linear (Hermitian) operators, they both must share the same eigenstates. Hence, $ D|f\rangle = d |f\rangle$, say. This leads to $F\left(H|f\rangle\right) = (d-f)\left(H|f\rangle\right)$ and $F\left(H^\dagger|f\rangle\right) = (d-f)\left(H^\dagger|f\rangle\right)$. And, this result matches exactly that of $\left\{H,H^\dagger\right\}=d\mathbb{I}$ case. We also observe that $ D|1-f\rangle = d |1-f\rangle$. This is consistent with the restriction to a single eigenvalue of $D$.

\section{Non-Hermitian Hilbert space on Bloch sphere}
It is known that the complete Hilbert space of a Hermitian Hamiltonian in $2D$ Hilbert space is constrained to lie on a line on the Bloch sphere (the orthogonal eigenstates lie on antipodal points of the sphere). But we observe from our Bloch sphere picture of the energy eigenstates in Fig.~$2$ of the main text that the complete Hilbert space of the non-Hermitian operator (including $|E_\pm\rangle$ and its dual $|\tilde{E}_\pm\rangle$) is constrained on a plane specified by the corresponding $\left(\theta, \phi \right)$. (It should be kept in mind that the states $|f\rangle$ and $|1-f\rangle$ themselves have their own Bloch sphere).

\section{Dual space }
In this section, we will find the unitary and anti-unitary operators connecting $|E_\pm\rangle$ and $|\tilde{E}_\pm\rangle$. This will eliminate the need of the two-step process, namely, quantizing both $H$ and $H^\dagger$, in specifying the well-defined dual space for non-Hermitian systems. In addition, we find that our analysis reveals a hidden $Z_2$ symmetry of the system which is also a metric in our theory.

\subsection{Unitary operators connecting $|E_\pm\rangle$ and $|\tilde{E}_\pm\rangle$}
We start with finding the unitary operator connecting $|E_\pm\rangle$ and $|\tilde{E}_\pm\rangle$. It can be achieved in two ways, (i) $|\tilde{E}_\pm\rangle  \propto  \mathcal{U}_1 |E_\pm\rangle$, and (ii) $|\tilde{E}_\pm\rangle  \propto  \mathcal{U}_2 |E_\mp\rangle$. We consider these two cases separately.

\subsubsection{Unitary operator $\mathcal{U}_1$,  $|\tilde{E}_\pm\rangle \propto \mathcal{U}_1 |E_\pm\rangle$}
We begin to find the unitary operator $\mathcal{U}_1$, $\mathcal{U}_1 |E_\pm\rangle = m_\pm |\tilde{E}_\pm\rangle$, where $m_\pm \in \mathbb{C}$. For the ease of notation in the following analysis, we first redefine the energy eigenstates and their biorthogonal partners  as $|E_\pm\rangle'=\sqrt{|a|}|E_\pm\rangle$ and $|\tilde{E}_\pm\rangle'=|\tilde{E}_\pm\rangle/\sqrt{|a|}$, i.e., by excluding the biorthogonal normalization factor of $\sqrt{|a|}$. This gives, $\mathcal{U}_1 |E_\pm\rangle' = m'_\pm |\tilde{E}_\pm\rangle'$, where $m'_\pm =|a|m_\pm$. From our expressions for $|E_\pm\rangle$ and $|\tilde{E}_\pm\rangle$ in the computational space, this gives,
\begin{equation}\label{Supp_u1}
    \mathcal{U}_1 |f\rangle \pm |a| e^{i\phi} \mathcal{U}_1 |1-f\rangle  = m'_\pm|f\rangle \pm \frac{e^{i\phi}}{|a|} m'_\pm |1-f\rangle.
\end{equation}

We express the action of $\mathcal{U}_1$ on the computational space as, 
\begin{subequations}\label{Supp_u1 hhs}
\begin{align}
    \mathcal{U}_1 |f\rangle &= c_1 |f\rangle +c_2|1-f\rangle,\\
    \mathcal{U}_1 |1-f\rangle &= d_1 |f\rangle +d_2|1-f\rangle.
\end{align}
\end{subequations}
where $c_i,d_i \in \mathbb{C}$ and $ i=1,2$. We note that $c_i, d_i$ completely specify $\mathcal{U}_1$ as the computational basis forms a complete Hilbert space. The unitariry of $\mathcal{U}_1$ imposes the following constraints on $c_i, d_i$,
\begin{subequations}\label{Supp_cidi}
\begin{align}
    |c_1|^2 + |c_2|^2 &= 1, \label{Supp_cidi a}\\
    |d_1|^2 + |d_2|^2 &= 1, \label{Supp_cidi b}\\
    d_1^* c_1 +d_2^* c_2 &=0,
\end{align}
\end{subequations}
as it leaves the Hermitian inner product invariant. Using Eq.~\eqref{Supp_u1 hhs} in \eqref{Supp_u1} gives,
\begin{align}
    \left( c_1 \pm |a| e^{i\phi} d_1 \right) |f\rangle +& \left( c_2 \pm |a| e^{i\phi} d_2 \right) |1-f\rangle \nonumber \\
    &= m'_\pm |f\rangle \pm \frac{e^{i\phi}}{|a|} m'_\pm |1-f\rangle.
\end{align}
Linear independence of $|f\rangle$ and $|1-f\rangle$ implies,
\begin{subequations}\label{Supp_cidimpm}
\begin{align}
    c_1 \pm |a| e^{i\phi} d_1 &= m'_\pm,\\
    c_2 \pm |a| e^{i\phi} d_2 &= \pm \frac{e^{i\phi}}{|a|} m'_\pm.
\end{align}
\end{subequations}

Using Eq.~\eqref{Supp_cidimpm} to express $c_i, d_i$ in terms of $m'_\pm$ as below,
\begin{subequations}\label{Supp_cidimpm2}
\begin{align}
    c_1 &= \frac{m'_+ +m'_-}{2}, \label{Supp_cidimpm2 a}\\
    c_2 &= \frac{e^{i\phi}}{|a|} \left(\frac{m'_+-m'_-}{2}\right), \label{Supp_cidimpm2 b}\\
    d_1 &= \frac{e^{-i\phi}}{|a|} \left(\frac{m'_+-m'_-}{2}\right),\\
    d_2 &= \frac{1}{|a|^2} \left(\frac{m'_+ +m'_-}{2}\right).
\end{align}
\end{subequations}

These four relations (Eq.~\eqref{Supp_cidimpm2}) reduce the number of unknown complex variables in Eq.~\eqref{Supp_u1 hhs} to two (as opposed to four that we started out with) as,
\begin{equation}\label{Supp_di rel ci}
    d_1 = e^{-2i\phi} c_2\, ,   \qquad d_2 =\frac{1}{|a|^2} c_1.
\end{equation}
Now, using equation \eqref{Supp_cidi b} with \eqref{Supp_di rel ci} we find, $\frac{1}{|a|^4} |c_1|^2 +|c_2|^2 =1$. Comparing this with Eq.~\eqref{Supp_cidi a}, we conclude that $c_1=d_2=0$ for $|a|\neq 1$. Note that $|a|=1$ corresponds to the NP. We discuss this case in detail in Appendix A2 of the main text.
Using this with Eq.~\eqref{Supp_cidimpm2 a}, we find, $ m'_+ = -m'_-= m'\,\textit{(say)}$. And, Eq.~\eqref{Supp_cidimpm2 b} and \eqref{Supp_di rel ci} gives, $c_2,d_1 = \frac{e^{\pm i\phi}}{|a|} m'$. Eq.~\eqref{Supp_cidi a} now implies $|m'| = |a|$. Let us call $m'= |a| e^{i\eta}$, where $\eta \in \mathbb{R}$. We find that,
\begin{subequations}\label{Supp_u1 hhs2}
\begin{align}
    \mathcal{U}_1 |f\rangle &= e^{i\left(\eta + \phi \right)} |1-f\rangle,\\
    \mathcal{U}_1 |1-f\rangle &= e^{i\left(\eta - \phi \right)} |f\rangle,
\end{align}
\end{subequations}
i.e., $\mathcal{U}_1$ acts as ladder operator (with cyclic ladder action) for the computational space. We note from Eq.~\eqref{Supp_u1 hhs2} that the only parameter that remains unknown in the specification $\mathcal{U}_1$ is $\eta$. For that, we observe that $e^{i\eta}$ appears as a global gauge in the action of $\mathcal{U}_1$ on the computational space. Consequently, we set $e^{i\eta}=1$. Finally, for the biorthogonal energy eigenspace, this gives,
\begin{equation}\label{Supp_u1 2}
    \mathcal{U}_1 |E_\pm\rangle = \pm  |\tilde{E}_\pm\rangle,
\end{equation}
i.e., $\mathcal{U}_1$ takes $|E_\pm\rangle$ to $|\tilde{E}_\pm\rangle$ with a relative phase difference of $\pi$. We observe in addition that $\mathcal{U}_1^{^2}|f\rangle = |f\rangle$, $\mathcal{U}_1^{^2}|f\rangle = |f\rangle$, $\mathcal{U}_1^{^2}|E_\pm\rangle = |E_\pm\rangle$ and $\mathcal{U}_1^{^2}|\tilde{E}_\pm\rangle = |\tilde{E}_\pm\rangle$, which implies $\mathcal{U}_1^{\prime^2} =\mathbb{I}$. Hence $\mathcal{U}_1$ is necessarily a discrete operator
\newline


\underline{\textit{Dual space and hidden $\mathbb{Z}_2$ symmetry of the system}:} We now begin to define the dual space of the non-Hermitian Hamiltonian that leads to orthogonal energy states with positive-definite norm. We first note from Eq.~\eqref{Supp_u1 2} that,
\begin{subequations}
    \begin{align}
        \langle \mathcal{U}_1 E_\pm |E_\pm\rangle  &= \pm 1,\\
        \langle \mathcal{U}_1 E_\pm |E_\mp\rangle  &= 0.
    \end{align}
\end{subequations}
Hence the unitary map $\mathcal{U}_1$ leads to negative norm states. We remedy this situation by defining the metric operator that compensates for this negative sign as $\mathcal{C}_1|E_\pm\rangle=\pm|E_\pm\rangle$ and leads to a positive definite inner product $\langle \mathcal{C}_1^\dagger \mathcal{U}_1 E_n | E_m\rangle  = \delta_{nm}$. Inspired from the charge conjugation operator in PT-symmetric quantum theory, such metric operator in the biorthogonal energy space can be expressed as below,
\begin{equation}
    \mathcal{C}_1 = |E_+\rangle \langle \mathcal{U}_1 E_+ | + |E_-\rangle \langle \mathcal{U}_1 E_-|,
\end{equation}
whereas in the computational basis it takes the following off-diagonal form,
\begin{equation}
    \mathcal{C}_1 = a  |1-f\rangle \langle f| +  \frac{1}{a} |f\rangle \langle 1-f|,
\end{equation}
where $a=|a|e^{i\phi}$. And, $\mathcal{C}_1$ is a non-Hermitian operator which becomes Hermitian only at the normal point ($|a|=1$). We note that since $\mathcal{C}_1$ and $H$ share the same eigenstates, $\mathcal{C}_1$ is a symmetry of the system which we call the 'hidden-symmetry' in analogy with the PT-quantum theory literature.

\subsubsection{Unitary operator $\mathcal{U}_2$,  $|\tilde{E}_\mp\rangle \propto \mathcal{U}_2 |E_\pm\rangle$}
We now consider $\mathcal{U}_2 |E_\pm\rangle = n_\pm |\tilde{E}_\mp\rangle$, $n_\pm \in \mathbb{C}$. With the redefined energy states, $\mathcal{U}'_2 |E_\pm\rangle' = n'_\pm |\tilde{E}_\mp\rangle'$ where $n'_\pm=|a|n_\pm$. We define the action of $\mathcal{U}_2$ on the computational as,
\begin{subequations}\label{Supp_u2 hhs}
\begin{align}
    \mathcal{U}_2 |f\rangle &= l_1 |f\rangle +l_2|1-f\rangle,\\
    \mathcal{U}_2 |1-f\rangle &= h_1 |f\rangle +h_2|1-f\rangle,
\end{align}
\end{subequations}
with $l_i, h_i \in \mathbb{C}$ and $i =1,2$, which leads to the following contraints due to the unitarity of $\mathcal{U}_2$. 
\begin{subequations}\label{Supp_lihi}
\begin{align}
    |l_1|^2 + |l_2|^2 &= 1, \label{Supp_lihi a}\\
    |h_1|^2 + |h_2|^2 &= 1, \label{Supp_lihi b}\\
    h_1^* l_1 +h_2^* l_2 &=0.
\end{align}
\end{subequations}
With the same procedure as followed for $\mathcal{U}_1$ in the previous section, we find,
\begin{equation}
    h_1 = -e^{-2i\phi} l_2\, ,\qquad
    h_2 = -\frac{1}{|a|^2} l_1,
\end{equation}
which gives $l_1 = h_2 =0$ and $n'_+ = -n'_-= n' \, \textit{(say)}$ for $|a| \neq 1$. This further reduces the number of unknown complex parameters in Eq.~\eqref{Supp_u2 hhs} as 
$l_2, h_1= \mp\frac{e^{\pm i\phi}}{|a|} n'$. Similar to the analysis for $\mathcal{U}_1$, we find $|n'| =|a|$ and we parameterize $n'$ as $n= |a| e^{i\lambda}$ where $\lambda \in \mathbb{R}$. For this case as well, we can set, $ e^{i\lambda}=1$since it appears as a global gauge. This specifies $\mathcal{U}_2$ through its action on the computational space as below,
\begin{subequations}\label{Supp_u2 hhs2}
\begin{align}
    \mathcal{U}_2 |f\rangle &=  - e^{i\phi} |1-f\rangle,\\
    \mathcal{U}_2 |1-f\rangle &= e^{-i\phi} |f\rangle.
\end{align}
\end{subequations}
Finally, $\mathcal{U}_2$ maps to the biorthogonal dual space as,
\begin{equation}\label{Supp_u2 nhhs}
    \mathcal{U}_2 |E_\pm\rangle = \pm  |\tilde{E}_\mp\rangle.
\end{equation}
Additionally, we note that $\mathcal{U}_2^{^2}|f\rangle = -|f\rangle$, $\mathcal{U}_2^{^2}|1-f\rangle = -|1-f\rangle$, $\mathcal{U}_2^{^2}|E_\pm\rangle = -|E_\pm\rangle$ and $\mathcal{U}_2^{^2}|\tilde{E}_\pm\rangle = -|\tilde{E}_\pm\rangle$. This implies that $\mathcal{U}_2^{\prime^2} - -\mathbb{I}$ and hence $\mathcal{U}_2$ is a discrete operator. We emphasize here that this overall phase in the square of the unitary operator is not gauge invariant (as opposed to that of an anti-unitary operator), and hence has no physical meaning. The negative sign here is just a reflection of our global gauge choice.


\underline{\textit{Dual space}:} Eq.~\eqref{Supp_u2 nhhs} implies, $\langle \mathcal{U}_2 E_\pm | E_\pm \rangle =0$ and $\langle \mathcal{U}_2^\prime E_\pm | E_\mp \rangle =\pm 1$. Similar to $\mathcal{U}_1$, we remedy the negative norm situation by defining the metric operator,
\begin{equation}
     \mathcal{C}_2 = |E_+\rangle \langle \mathcal{U}_2 E_- | + |E_-\rangle \langle \mathcal{U}_2 E_+|,
\end{equation}
which compensates the minus sign since $\mathcal{C}_2 |E_\pm\rangle = \mp  |E_\pm\rangle$. The positive definite inner product can now be defined as $\langle \mathcal{C}_2^{\dagger} \mathcal{U}_2 E_n | E_m\rangle  = \left(1-\delta_{nm}\right)$. By expressing $\mathcal{C}_2$ in the computation basis, we observe that $\mathcal{C}_2=-\mathcal{C}_1$. Hence, $\mathcal{C}_2$ does not lead to a new hidden-symmetry of the system.

\subsection{Anti-unitary operators connecting $|E_\pm\rangle$ and $|\tilde{E}_\pm\rangle$}
Similar to the unitary maps, we have two cases here, (i) $|\tilde{E}_\pm\rangle  \propto  \mathcal{A}_1 |E_\pm\rangle$, and (ii) $|\tilde{E}_\pm\rangle  \propto  \mathcal{A}_2 |E_\mp\rangle$.

\subsubsection{Anti-unitary operator $\mathcal{A}_1$,  $|\tilde{E}_\pm\rangle \propto \mathcal{A}_1 |E_\pm\rangle$}
We have, $\mathcal{A}_1 |E_\pm\rangle = p_\pm |\tilde{E}_\pm\rangle$ where $p_\pm \in \mathbb{C}$. In the redefined basis,  $\mathcal{A}'_1 |E_\pm\rangle' = p'_\pm |\tilde{E}_\pm\rangle'$ where $ p'_\pm=|a|  p_\pm$. Using the expansion of $|E_\pm\rangle$ and $|\tilde{E}_\pm\rangle$ in the computational basis, we have,
\begin{equation}\label{Supp_a1}
    \mathcal{A}_1\left( |f\rangle \pm |a| e^{i\phi} |1-f\rangle \right) = p'_\pm \left(|f\rangle \pm \frac{e^{i\phi}}{|a|}  |1-f\rangle \right).
\end{equation}
Using anti-unitary property of $\mathcal{A}_1$, this gives,
\begin{equation}\label{Supp_a1 2}
    \mathcal{A}_1 |f\rangle \pm |a| e^{-i\phi} \mathcal{A}_1 |1-f\rangle  = p'_\pm|f\rangle \pm \frac{e^{i\phi}}{|a|} p'_\pm |1-f\rangle.
\end{equation}
We first define the action of $\mathcal{A}_1$ on the computational basis as below,
\begin{subequations}\label{Supp_a1 hhs}
\begin{align}
    \mathcal{A}_1 |f\rangle &= r_1 |f\rangle +r_2|1-f\rangle,\\
    \mathcal{A}_1 |1-f\rangle &= s_1 |f\rangle +s_2|1-f\rangle.
\end{align}
\end{subequations}
where $r_i,s_i \in \mathbb{C}$ and $ i=1,2$. We now note that the action of anti-unitary leaves orthonormality of the states invariant. This imposes the following constraints,
\begin{subequations}\label{Supp_risi}
\begin{align}
    |r_1|^2 + |r_2|^2 &= 1, \label{Supp_risi a}\\
    |s_1|^2 + |s_2|^2 &= 1, \label{Supp_risi b}\\
    s_1^* r_1 +s_2^* r_2 &=0.
\end{align}
\end{subequations}

Proceeding similarly as for $\mathcal{U}_{1,2}$, we find,
\begin{equation}
    s_1 = r_2\, , \qquad s_2 = \frac{e^{2i\phi}}{|a|^2} r_1,
\end{equation}
which gives $r_1 =s_2 =0$, $p'_+ = -p'_-= p' \, \textit{(say)}$ and $|p'| = |a|$ for $|a| \neq 1$. We parameterize $p'$ as $p'=|a| e^{i\mu}$,  where $\mu \in \mathbb{R}$, and set $e^{i\mu}=1$ for precisely the same reason as explained for $\mathcal{U}_{1,2}$. This specifies $\mathcal{A}_1$ through its action on the computational space as below,
\begin{subequations}\label{Supp_a1 hhs2}
\begin{align}
    \mathcal{A}_1 |f\rangle &= e^{i\phi} |1-f\rangle,\\
    \mathcal{A}_1 |1-f\rangle &= e^{i\phi} |f\rangle.
\end{align}
\end{subequations}
Since, $e^{i\phi}$ also appears as a global phase in the action of $\mathcal{A}_1$ on the computational space, we absorb this as well in the definition of $\mathcal{A}_1$. For the energy eigenspace this implies,
\begin{equation}\label{Supp_a1 2}
    \mathcal{A}_1 |E_\pm\rangle = \pm e^{-i\phi}   |\tilde{E}_\pm\rangle.
\end{equation}
 We additionally note that $\mathcal{A}_1^{2}|f\rangle = |f\rangle$, $\mathcal{A}_1^{2}|1-f\rangle = |1-f\rangle$, $\mathcal{A}_1^{2}|E_\pm\rangle = |E_\pm\rangle$ and $\mathcal{A}_1^{2}|\tilde{E}_\pm\rangle = |\tilde{E}_\pm\rangle$. This implies $\mathcal{A}_1^{2} =  \mathbb{I}$ (where the phase on the right hand side is gauge invariant).
\newline


\underline{\textit{Dual space}:} Again, Eq.~\eqref{Supp_a1 2}, leads to negative norm states. We deal with this situation by defining the metric operator as below,
\begin{equation}
     \mathcal{C}_3 = e^{-i\phi}\left(|E_+\rangle \langle \mathcal{A}_1 E_+ | + |E_-\rangle \langle \mathcal{A}_1 E_-|\right),
\end{equation}
which compensates the minus sign, $\mathcal{C}_3 |E_\pm\rangle = \pm e^{-i\phi}|E_\pm\rangle$, and leads to positive-definite norm as $\langle \mathcal{C}_3^{\dagger} \mathcal{A}_1 E_n | E_m\rangle  = \delta_{nm}$. Expressing $\mathcal{C}_3$ in the computational basis, we find, $\mathcal{C}_3=e^{-i\phi}\mathcal{C}_1$.

\subsubsection{Anti-unitary operator $\mathcal{A}_2$,  $|\tilde{E}_\mp\rangle \propto \mathcal{A}_2 |E_\pm\rangle$}
Next, we consider $\mathcal{A}_2 |E_\pm\rangle = q_\pm |\tilde{E}_\mp\rangle$ where $q_\pm \in \mathbb{C}$. With $\mathcal{A}'_2 |E_\pm\rangle' = q'_\pm |\tilde{E}_\mp\rangle'$ where $ q'_\pm=|a|  q_\pm$. We define action of $\mathcal{A}_2$ on the computational space as.
\begin{subequations}\label{Supp_a2 hhs}
\begin{align}
    \mathcal{A}_2 |f\rangle &= v_1 |f\rangle +v_2|1-f\rangle,\\
    \mathcal{A}_2 |1-f\rangle &= w_1 |f\rangle +w_2|1-f\rangle,
\end{align}
\end{subequations}
with $v_i, w_i \in \mathbb{C}$ and $i =1,2$. The anti-unitarity of $\mathcal{A}_2$ imposes the following constraints,
\begin{subequations}\label{Supp_viwi}
\begin{align}
    |v_1|^2 + |v_2|^2 &= 1, \label{Supp_viwi a}\\
    |w_1|^2 + |w_2|^2 &= 1, \label{Supp_viwi b}\\
    w_1^* v_1 +w_2^* v_2 &=0.
\end{align}
\end{subequations}
We find, as before, $w_1 = -v_2$, $ w_2 = -\frac{e^{2i\phi}}{|a|^2} v_1$ which implies $v_1=w_2=0$ for $|a| \neq 1$. This gives, $q'_+ = -q'_-= q' \, \textit{(say)}$ and and $|q'| = |a|$. We parameterize $q'$ as $q'=|a|e^{i\mu}$ and set $e^{i\mu}=1$ for similar reasons as before. $\mathcal{A}_2$ is now specified as,
\begin{subequations}\label{Supp_a2 hhs2}
\begin{align}
    \mathcal{A}_2 |f\rangle &=  - e^{i\phi} |1-f\rangle,\\
    \mathcal{A}_2 |1-f\rangle &= e^{i\phi} |f\rangle.
\end{align}
\end{subequations}
After absorbing $e^{i\phi}$ in the definition of $\mathcal{A}_2$, this transforms the energy eigenspace as,
\begin{equation}\label{Supp_a2 2}
    \mathcal{A}_2 |E_\pm\rangle = \pm e^{-i\phi}  |\tilde{E}_\mp\rangle.
\end{equation}
Next, we note that $\mathcal{A}_2^{2}|f\rangle = -|f\rangle$, $\mathcal{A}_2^{2}|1-f\rangle = -|1-f\rangle$, $\mathcal{A}_2^{2}|E_\pm\rangle = -|E_\pm\rangle$ and $\mathcal{A}_2^{2}|\tilde{E}_\pm\rangle = -|\tilde{E}_\pm\rangle$, which implies $\mathcal{A}_2^{^2} = -\mathbb{I}$. The minus sign here is gauge invariant.
\newline


\underline{\textit{Dual space}:} To deal with the negative norm situtation arising from Eq.~\eqref{Supp_a2 2}, we define the metric operator, 
\begin{equation}
     \mathcal{C}_4 = e^{-i\phi}\left(|E_+\rangle \langle \mathcal{A}_2 E_- | + |E_-\rangle \langle \mathcal{A}_2 E_+|\right),
\end{equation}
which acts on the energy space as $\mathcal{C}_4 |E_\pm\rangle = \mp e^{-i\phi} |E_\pm\rangle$ and leads to a postive definite norms as $\langle \mathcal{C}_4^{\dagger} \mathcal{A}_2 E_m | E_n\rangle  = \left(1-\delta_{nm}\right)$. By expressing $\mathcal{C}_4$ in the computational basis, we observe that $\mathcal{C}_4=-\mathcal{C}_3$.

\section{Some comments on circular and point degeneracy}
We note here some important features of the circular degeneracy,
\begin{enumerate}
    \item We observe from Eq. (23) in the main text that $\langle E_{\pm}^n|E_{\pm}^m\rangle =\left(1+|a|^2\right)\delta_{mn}$ and $\langle E_{\mp}^n|E_{\pm}^m\rangle =\left(1-|a|^2\right)\delta_{mn}$. Hence, the states $|E_+^n\rangle$ are orthogonal amongst themselves (same for $|E_-^n\rangle$). And, the only overlapping states are the particle-hole partners $|E_\pm^n\rangle$.

    \item For a $2N-$dimensional circularly degenerate subspace, this suggests the existence of $N-$EP2's (collapse of $N-$ pairs of particle-hole partners $|E_\pm^n\rangle$) at the boundary of the computational spectrum $f=0,1\,$. Hence, the subspace dimensionality reduces to $n$ (half of its original dimesionality). We also note that at this exceptional contour, $H$ annihilates the zero eigenvalue computational subspace and acts as a lowering operator for the states in the  computational subspace corresponding to the eigenvalue $1$.
\end{enumerate}
These observations can be appropriately extended to the point degeneracy case.

\section{$\left(F-\frac{\mathbb{I}}{2}\right)$ as ladder operator for energy eigenstates}
Next, we note that just as $H$ and $H^\dagger$ act as ladder operators for the computational basis, a suitable rescaling of computational spectrum leads to a ladder operator for the eigenstates of $H$ and $H^\dagger$. We set $d=1$ in this section and parameterize $f$ as $f=\frac{1}{2}+|\mathfrak{f}|$ where $0 \leq |\mathfrak{f}| \leq \frac{1}{2}$. This gives,
\begin{subequations}
\begin{align}
    \left(F-\mathbb{I}/2 \right) |f\rangle &=  |\mathfrak{f}| |f\rangle,\\
    \left(F-\mathbb{I}/2 \right) |1-f\rangle &= - |\mathfrak{f}| |1-f\rangle,
\end{align}
\end{subequations}
which leads to $\left(F-\mathbb{I}/2 \right)|E_\pm\rangle=|\mathfrak{f}||E_\mp\rangle$ and $\left(F-\mathbb{I}/2 \right)|\tilde{E}_\pm\rangle=|\mathfrak{f}||\tilde{E}_\mp\rangle$. Hence, we observe that $\left(F-\mathbb{I}/2 \right)$ acts as ladder operator for the energy space and its biorthogonal space. We note here that this ladder action projects to the respective states with the same prefactor $|\mathfrak{f}|$.

\section{An example in $2D$-Hilbert space (PT-symmetry as a special case in our formalism)}
We consider a $2 \times 2$ Hamiltonian, $h_x=s,\, h_y=0, h_z=r_1+ir_2$ with $r_{1,2},s \in \mathbb{R}$ and $\left\{H,H^\dag\right\}=2\left(r_1^2+r_2^2+s^2\right)\mathbb{I}$. We set $2\left(r_1^2+r_2^2+s^2\right)=1$. The corresponding Hermitian operator $F$ is found to be $\mathfrak{f}_x=\mathfrak{f}_z=0,\, \mathfrak{f}_y=2r_2s$. The eigenvalues of $F$ are $f=r_1^2 + \left(s-r_2\right)^2$ and $1-f=r_1^2 + \left(s+r_2\right)^2$ with the computational basis states being the eigenstates of the $\sigma_y$ operator, \begin{align}
\left|f\right\rangle =\frac{1}{\sqrt{2}}\left(\begin{array}{c}
1 \\
-i
\end{array}\right),\quad 
\left|1-f\right\rangle =\frac{1}{\sqrt{2}}\left(\begin{array}{l}
1 \\
i
\end{array}\right).
\end{align}
Off-block diagonalization of $H$ gives $A=e^{i(\gamma+\phi)} \sqrt{f}=r_1+i(r_2-s)$ and $B=e^{i(\gamma-\phi)} \sqrt{1-f}=r_1+i(r_2+s)$. The energy eigenvalues are $E_\pm=\pm\sqrt{AB}=\pm\sqrt{(r_1+ir_2)^2+s^2}$ and the energy eigenvectors are $|E_\pm\rangle = \frac{1}{\sqrt{2}}\left(|f\rangle \pm \left(\frac{A}{E}\right) |1-f\rangle \right)$ where $E$ in the expansion coefficient is defined as $E_\pm = \pm E$ and $a=\frac{A}{E}$. 
The two phases can be found as $\gamma=\textit{arg}(E)$ and $\phi=\textit{arg}(\frac{A}{E})$.

The transformation to the dual space can be made using $\mathcal{U}_i$ or $\mathcal{A}_i$. Here we demonstrate it for the antiunitary. Take $\mathcal{A}_1=\mathcal{V}_1\mathcal{K}$, for example. In the computational basis, $\mathcal{V}_1$ acts as $\sigma_x$ and $\mathcal{K}$ leaves the computational basis states invariant. This gives,
\begin{align}
    \mathcal{A}_1|E_\pm\rangle &= \frac{1}{\sqrt{2}}\left(|1-f\rangle \pm \left(\frac{A^*}{E^*}\right) |f\rangle \right) \nonumber \\
    &= \pm \frac{a^*}{\sqrt{2}} \left(|f\rangle \pm \left(\frac{1}{a^*}\right) |1-f\rangle \right) = \pm a^* |\tilde{E}_\pm\rangle.
\end{align}
Similarly, $\mathcal{A}_2|E_\pm\rangle= \pm a^* |\tilde{E}_\mp\rangle$.

Next, we note that the projection of the above Hamiltonian to the $r_1\rightarrow 0$ plane exhibits PT-symmetry with $\mathcal{P}=\tau_x$ and $\mathcal{T}=\mathcal{K}$ in the Hamiltonian basis. We now study this projected Hamiltonian to see PT-symmetry as a special case in our formalism. From the expressions of $A$ and $B$, we find, $\left(\phi\pm\gamma\right)=\textit{tan}^{-1}\left(\frac{r_2\mp s}{r_1}\right)$. For $|s|>|r_2|$ and $s\rightarrow +ve$ ($s\rightarrow -ve$), this leads to $\left(\phi+\gamma\right)=\frac{3\pi}{2}\left(\frac{\pi}{2}\right)$ and $\left(\phi-\gamma\right)=\frac{\pi}{2}\left(\frac{3\pi}{2}\right)$. Also, for $|s|<|r_2|$ and $r_2\rightarrow +ve$ ($r_2\rightarrow -ve$), we find $\left(\phi+\gamma\right)=\frac{\pi}{2}\left(\frac{3\pi}{2}\right)$ and $\left(\phi-\gamma\right)=\frac{\pi}{2}\left(\frac{3\pi}{2}\right)$. This leads to real eigenvalues in $|s|>|r_2|$ which matches with the PT-theory results.



\end{document}